\documentclass[12pt]{article}
\usepackage{amsmath,amssymb,amscd,amsthm,color,graphics,verbatim}
\usepackage[margin=1.1in]{geometry}

\usepackage[enableskew]{youngtab}
\def\qed{\hfill $\vrule height 2.5mm  width 2.5mm depth 0mm $}

\newtheorem{theorem}{Theorem}
\newtheorem{proposition}[theorem]{Proposition}

\newtheorem{corollary}[theorem]{Corollary}
\newtheorem{conjecture}[theorem]{Conjecture}

\theoremstyle{definition}

\newtheorem{definition}[theorem]{Definition}
\newtheorem{remark}[theorem]{Remark}
\newtheorem{example}[theorem]{Example}

\begin{document}
$\,$\vspace{0mm}

\begin{center}
{\sf\huge Singular Solutions to the Bethe Ansatz}
\vspace{3mm}\\
{\sf\huge Equations and Rigged Configurations}
\vspace{15mm}\\
{\textsf{\LARGE  ${}^{\mbox{\small a}}$Anatol N. Kirillov and
${}^{\mbox{\small b}}$Reiho Sakamoto}}
\vspace{20mm}\\
{\textsf {${}^{\mbox{\small{a}}}$Research Institute for Mathematical Sciences,}}
\vspace{-1mm}\\
{\textsf {Kyoto University, Sakyo-ku,}}
\vspace{-1mm}\\
{\textsf {Kyoto, 606-8502, Japan}}
\vspace{-1mm}\\
{\textsf {kirillov@kurims.kyoto-u.ac.jp}}
\vspace{3mm}\\
{\textsf {${}^{\mbox{\small{b}}}$Department of Physics,}}
\vspace{-1mm}\\
{\textsf {Tokyo University of Science, Kagurazaka,}}
\vspace{-1mm}\\
{\textsf {Shinjuku, Tokyo, 162-8601, Japan}}
\vspace{-1mm}\\
{\textsf {reiho@rs.tus.ac.jp}}
\vspace{-1mm}\\
{\textsf {reihosan@08.alumni.u-tokyo.ac.jp}}
\vspace{30mm}
\end{center}

\begin{abstract}
\noindent
We provide a conjecture for the following two quantities
related with the spin-$\frac{1}{2}$ isotropic Heisenberg model
defined over rings of even lengths:
(i) the number of the solutions to the Bethe ansatz equations
which correspond to non-zero Bethe vectors;
(ii) the number of physical singular solutions of the Bethe ansatz equations
in the sense of Nepomechie--Wang.
The conjecture is based on a natural relationship between the solutions
to the Bethe ansatz equations and the rigged configurations.
\end{abstract}

\pagebreak

\section{Introduction}
The problem of constructing ``physical states" to
the isotropic Heisenberg model on a ring
(see Section \ref{sec:ABA} for the definition)
by the so-called Bethe ansatz method has been extensively studied
in both physical and mathematical literatures for more than 80 years 
after the seminal work \cite{Bethe} by Hans Bethe published in 1931.
There exists enormous number of papers concerning this problem,
but even for the simplest case of the $\mathfrak{sl}_2$ spin-$\frac{1}{2}$
isotropic Heisenberg model, it is widely regarded that there are
still remaining unclear aspects about the problem.
The goal of the present paper is to draw attention to a mysterious
connection between  ``{\it physical solutions}" of the Bethe ansatz equations
and combinatorial objects called the {\it  rigged configurations}.

For our purpose, we find it is convenient to use the so-called
algebraic Bethe ansatz method introduced by Faddeev's school
(\cite{FT}, see also \cite{{KorepinBook}}).
Basic procedure is as follows (see Section \ref{sec:ABA} for details).
We start from the state $|0\rangle$ in which all spins pointing up.
The state $|0\rangle$ is the obvious eigenvector of the Hamiltonian.
Then we construct a certain creation operator $B(\lambda)$ depending on
a parameter $\lambda\in\mathbb{C}$ to construct the other states
as $B(\lambda_1)\cdots B(\lambda_\ell)|0\rangle$ which we call the Bethe vectors.
The main observation is that if the parameters $\lambda,\ldots,\lambda_\ell$
satisfy the Bethe ansatz equations (see equation (\ref{eq:Bethe_ansatz}) below),
then the corresponding Bethe vector (if non-zero) is an eigenvector of the Hamiltonian.

However it is well known that the Bethe ansatz equations admit too many solutions
and many solutions correspond to the zero Bethe vector
(see \cite{HaoNepomechieSommese} and references therein).
Therefore one can state the main problem as follows:
Describe a set of solutions to the Bethe ansatz equations
which provide non-zero Bethe vectors after certain regularizations if necessary.

There is a common assumption such that it is enough to
consider solutions with pairwise distinct components.
Although there are no rigorous proofs of the assumption,
in the physical literature usually it is motivated by the {\it Pauli exclusion principle}.
However even if we impose this assumption, still there is a very subtle problem.
Indeed, if we discard all the solutions corresponding to the zero Bethe vector,
then some of the eigenvectors are missing from the Bethe vectors.

Let us call the solutions of the Bethe ansatz equations corresponding
to the zero Bethe vector the singular solutions
and those corresponding to the non-zero Bethe vectors the regular solutions.
Then our task is to find a set of singular solutions which provide
non-zero Bethe vectors after certain regularizations.
There are a large number of works which try to solve this problem
(see for example \cite{EKS:1992,Siddharthan,Beisert}).
Among them we are interested in a method to introduce
a higher order correction to the Bethe vectors.
In particular, recently Nepomechie--Wang \cite{NepomechieWang2013}
proposed an explicit criterion under which one can pick
all the missing physical states from the singular solutions.
Following \cite{NepomechieWang2013}
we call such solution {\it physical singular solution}.
Their conjecture is verified up to length 14 systems by
an extensive numerical computation \cite{HaoNepomechieSommese}.

The main observation of the present paper is that a combination
of both the regular solutions and the physical singular solutions has
a mysterious but natural correspondence with the combinatorial objects called
the rigged configurations.
The rigged configurations have been introduced by A. N. Kirillov and N. Reshetikhin
\cite{KirillovReshetikhin} (see also \cite{Kir1}, \cite{Kir4}, \cite{KSS:2002})
as an application of the so-called {\it string conjecture}~ \cite{Bethe,Takahashi}.
They have a canonical bijection with the tensor products of crystal bases
(see \cite{S:web} for a Mathematica implementation)
and known to possess deep structures related with subtle properties
of finite dimensional representations of the quantum affine algebras.
Moreover, they provide a complete set of the action-angle variables
of the box-ball system which is a prototypical example of
ultradiscrete (or tropical) soliton systems \cite{KOSTY}.

Motivated by the correspondence between the rigged configurations
and the solutions to the Bethe ansatz equations,
we propose a conjecture for the total number of the physical
solutions when the system size $N$ is even (see Conjecture \ref{conj:main}).
Since the total number of eigenvectors is well-known,
the conjecture in turn provides a total number of the regular solutions.
Our conjecture have a perfect agreement with the numerical data
provided in \cite{HaoNepomechieSommese}.

We remark that another method to construct Bethe vectors
corresponding to the so-called {\it admissible solutions} to the Bethe ansatz equations
has been developed in \cite{MTV}.

The organization of the present paper is as follows.
In Section \ref{sec:ABA}, we provide a necessary background about the
algebraic Bethe ansatz as well as a description of the Nepomechie--Wang's prescription.
In Section \ref{sec:RC}, we provide the definition of the rigged configurations
and state our main results.
In Section \ref{sec:discussion} we discuss a subtlety which appears
when the system size is odd.

 In the case when the  number of sites $N$ and the spin $\ell$  for the  spin-$\frac{1}{2}$ isotropic 
Heisenberg model are both \underline{even}, we  state conjectural formulas for the number of solutions with 
pairwise distinct roots, the number of singular solutions and the number of physical singular 
solutions, denoted respectively by ${\cal{N}},~{\cal{N}}_{s}$ and ${\cal{N}}_{sp}$ in 
\cite{HaoNepomechieSommese}.

\section{Algebraic Bethe ansatz analysis}
\label{sec:ABA}
The space of states $\mathfrak{H}_N$ and the Hamiltonian $\mathcal{H}_N$
of the spin-$\frac{1}{2}$ isotropic Heisenberg model on a length $N$ chain with
the periodic boundary condition are
\begin{align}
\mathfrak{H}_N &= \bigotimes_{j=1}^{N} V_j,\quad V_j \simeq {\mathbb {C}}^2,
\nonumber\\
\mathcal{H}_N &=  \frac{J}{4} \sum_{k=1}^{N}( \sigma_{k}^{x} \sigma_{k+1}^{x}
+ \sigma_{k}^{y} \sigma_{k+1}^{y}+ \sigma_{k}^{z} \sigma_{k+1}^{z} -{\mathbb {I}}_N),\qquad
\sigma_{N+1}^{a}=\sigma_{1}^{a}.
\end{align}
Here $\sigma^{a}$ $(a=x,y,z)$ are the Pauli matrices
\begin{align}
\sigma^{x} = 
\left(\!
\begin{array}{cc}
0&1\\
1&0
\end{array}
\!\right),\qquad
\sigma^{y} = 
\left(\!
\begin{array}{cc}
0&-i\\
i&0
\end{array}
\!\right),\qquad
\sigma^{z} = 
\left(\!
\begin{array}{cc}
1&0\\
0&-1
\end{array}
\!\right),
\end{align}
and the operators $\sigma_k^{a}$ $(a=x,y,z)$ act on $\mathfrak{H}_N$ as
\begin{align}
\sigma_{k}^{a}= I \otimes \cdots \otimes
\underbrace{\sigma^{a}}_{k}
\otimes \cdots \otimes I,
\end{align}
that is, they act non trivially only on the space $V_k$.
Here $I$ is the $2\times 2$ identity matrix and $\mathbb{I}_N$
is the identity operator on the space of states; $\mathbb{I}_N=I^{\otimes N}$.

Our task is to diagonalize the Hamiltonian exactly.
Instead of following the original arguments by Bethe,
we use the formalism called the algebraic Bethe ansatz \cite{KorepinBook}.
Let us denote the canonical basis vectors of $\mathbb{C}^2$ as
\begin{align}
v_+=
\left(\!
\begin{array}{c}
1\\0
\end{array}
\!\right),\qquad
v_-=
\left(\!
\begin{array}{c}
0\\1
\end{array}
\!\right).
\end{align}
Then the vector
\begin{align}
|0\rangle_N=v_+\otimes\cdots\otimes v_+
\in\mathfrak{H}_N
\end{align}
is an eigenvector of the Hamiltonian.
The basic idea of the algebraic Bethe ansatz is to construct
the remaining eigenvectors by using certain creation operators
$B(\lambda)$ $(\lambda\in\mathbb{C})$ acting on the state $|0\rangle_N$;
\begin{align}
\Psi_N(\lambda_1,\ldots,\lambda_\ell)
:=B_N(\lambda_1)\cdots B_N(\lambda_\ell)|0\rangle_N.
\end{align}
We call such vectors the Bethe vectors.
The definition of the operators $B_N(\lambda)$ is as follows.
We introduce $(2\times 2)$-size matrix operator $L_k(\lambda)$
\begin{align}
\label{eq:def_L_of_Bethe}
L_{k}(\lambda)=
\left(\!
\begin{array}{cc}
\lambda \mathbb{I}_N+\frac{i}{2}\sigma^z_k & \frac{i}{2}\sigma^-_k\\
\frac{i}{2}\sigma^+_k & \lambda \mathbb{I}_N-\frac{i}{2}\sigma^z_k
\end{array}
\!\right),
\end{align}
where $\sigma^\pm_k=\sigma^x_k\pm i\sigma^y_k$.
The operator $L_{k}(\lambda)$ acts on $\mathbb{C}^2\otimes\mathfrak{H}_N$
where $\mathbb{C}^2$ is an auxiliary space
representing 2 by 2 matrix in (\ref{eq:def_L_of_Bethe})
and operators like $\sigma^z_k$ act on $\mathfrak{H}_N$.
Then define the transfer matrix $T_N(\lambda)$ by
\begin{align}
T_N(\lambda)=L_N(\lambda)L_{N-1}(\lambda)\cdots L_1(\lambda)
\end{align}
and define the operators $A_N(\lambda),B_N(\lambda),C_N(\lambda)$ and $D_N(\lambda)$ by
\begin{align}
T_N(\lambda)=
\left(
\begin{array}{cc}
A_N(\lambda) & B_N(\lambda)\\
C_N(\lambda) & D_N(\lambda)
\end{array}
\right).
\end{align}
Since we can show that
$[B_N(\lambda_1),B_N(\lambda_2)]=0$,
we only need the set $\{\lambda_1,\ldots,\lambda_\ell\}$ modulo permutations
to specify the Bethe vector.

The fundamental observation is as follows.
The non-zero Bethe vector
\begin{align}
\Psi_N(\lambda_1,\ldots,\lambda_\ell)
=B_N(\lambda_1)\cdots B_N(\lambda_\ell)|0\rangle_N
\end{align}
is an eigenvector of the Hamiltonian
if and only if the numbers $\lambda_1,\ldots,\lambda_\ell$
satisfy the following system of algebraic equations
\begin{align}
\label{eq:Bethe_ansatz}
\left(
\frac{\lambda_k+\frac{i}{2}}{\lambda_k-\frac{i}{2}}
\right)^N
=\prod_{j=1 \atop j\neq k}^\ell
\frac{\lambda_k-\lambda_j+i}{\lambda_k-\lambda_j-i},
\qquad
(k=1,\cdots,\ell).
\end{align}
These equations are the celebrated Bethe ansatz equations
which we denote by $\mathrm{BAE}(N,\ell)$.
In the following, we only consider the solutions
$\{\lambda_1,\ldots,\lambda_\ell\}$ to the Bethe ansatz equations
which have pairwise distinct components; $\lambda_i\neq\lambda_j$ if $i\neq j$.
If a solution $\{\lambda_1,\ldots,\lambda_\ell\}$ of the Bethe ansatz equations
corresponds to a non-zero Bethe vector, we call such solution {\bf regular}.

Recall that the space of the states $\mathfrak{H}_N$
have natural action of the Lie algebra $\mathfrak{sl}_2$
such that
\begin{align}
S^a:=\frac{1}{2}\sum_{k=1}^N\sigma_k^a,\qquad (a=x,y,z).
\end{align}
Then we can show that the model possesses the $\mathfrak{sl}_2$ symmetry
$[\mathcal{H}_N,S^a]=0$.
Hence we can simultaneously diagonalize the Hamiltonian $\mathcal{H}_N$
and the operators $S^z$ and $S^2=(S^x)^2+(S^y)^2+(S^z)^2$.
Moreover, we can show that the Bethe vectors are highest weight vectors
\begin{align}
S^+\Psi_N(\lambda_1,\ldots,\lambda_\ell)=0,\qquad
S^\pm=S^x\pm iS^y
\end{align}
if the parameters $\lambda_1,\ldots,\lambda_\ell$
satisfy the Bethe ansatz equations (\ref{eq:Bethe_ansatz}).
In this case, the vector $\Psi_N(\lambda_1,\ldots,\lambda_\ell)$
is the eigenvector of $S^z$ with the eigenvalue $\frac{N}{2}-\ell$.
Therefore we can obtain the other vectors by acting the lowering operator $S^-$
successively on the Bethe vectors.
We denote by $\mathbf{m}$ the irreducible $\mathfrak{sl}_2$-module of dimension $m$.

\begin{remark}
Therefore it is enough to consider the case $\ell\leq\frac{N}{2}$.
However it is interesting to note the following conjecture.
If $\ell>\frac{N}{2}$, then for any solution $\lambda_1,\ldots,\lambda_\ell$
to the corresponding Bethe ansatz equations,
we have $B_N(\lambda_1)\cdots B_N(\lambda_\ell)|0\rangle_N=0$.
If we use the precise form of the operators $B_N(\lambda)$
we see that the corresponding vector vanishes non-trivially.
\qed
\end{remark}

However it is well known that there is a very subtle problem about the procedure.
Indeed, as already Bethe himself realized, the number of pairwise distinct solutions
to the Bethe ansatz equations is too large than the actual number of the eigenvectors,
and also some eigenvectors are missing from the Bethe vectors.
\begin{example}
Consider the case $N=4$.
We use the lexicographic ordering for the 16 basis vectors of $\mathfrak{H}_4$;
$|++++\rangle$, $|+++-\rangle$, $|++-+\rangle$, $|++--\rangle$, $\cdots$,
where we have used an abbreviated notation such as
\[
|+++-\rangle=v_+\otimes v_+\otimes v_+\otimes v_-.
\]

If $\ell=0$, the vector $|0\rangle_4$ provides the highest weight vector of the representation $\mathbf{5}$.
If $\ell=1$, the Bethe ansatz equation
\begin{align*}
\left(
\frac{\lambda_1+\frac{i}{2}}{\lambda_1-\frac{i}{2}}
\right)^4
=1,
\end{align*}
has the solutions $\lambda_1=0,\pm\frac{1}{2}$.
Then we have
\begin{align*}
8B_4(0)|0\rangle_4&=(0, -1, 1, 0, -1, 0, 0, 0, 1, 0, 0, 0, 0, 0, 0, 0)^t,\\
\frac{4}{1-i}B_4\!\left(\frac{1}{2}\right)|0\rangle_4&=(0, 1, i, 0, -1, 0, 0, 0, -i, 0, 0, 0, 0, 0, 0, 0)^t,\\
\frac{4}{1+i}B_4\!\left(-\frac{1}{2}\right)|0\rangle_4&=(0, 1, -i, 0, -1, 0, 0, 0, i, 0, 0, 0, 0, 0, 0, 0)^t.
\end{align*}
By suitable combinations of these vectors, we have three highest weight vectors
corresponding to $\mathbf{3}^{\oplus 3}$.

Let us consider the case $\ell=2$.
Then the Bethe ansatz equations for this case is
\begin{align*}
\left(
\frac{\lambda_1+\frac{i}{2}}{\lambda_1-\frac{i}{2}}
\right)^4
=
\frac{\lambda_1-\lambda_2+i}{\lambda_1-\lambda_2-i},\qquad
\left(
\frac{\lambda_2+\frac{i}{2}}{\lambda_2-\frac{i}{2}}
\right)^4
=
\frac{\lambda_2-\lambda_1+i}{\lambda_2-\lambda_1-i}.
\end{align*}
If we assume that all $\lambda_j$ are distinct, we have the following four solutions;
\[
\{\lambda_1,\lambda_2\}=
\{\frac{i}{2},-\frac{i}{2}\},\,
\{-\frac{i}{2},\frac{i}{2}\},\,
\{\frac{1}{\sqrt{12}},-\frac{1}{\sqrt{12}}\},\,
\{-\frac{1}{\sqrt{12}},\frac{1}{\sqrt{12}}\}.
\]
We can confirm that
\[
B_4\!\left(\frac{i}{2}\right)B_4\!\left(-\frac{i}{2}\right)=
B_4\!\left(-\frac{i}{2}\right)B_4\!\left(\frac{i}{2}\right)=0.
\]
On the other hand, we have
\[
\frac{27}{2}B_4\!\left(\frac{1}{\sqrt{12}}\right)B_4\!\left(-\frac{1}{\sqrt{12}}\right)|0\rangle_4=
(0, 0, 0, 1, 0, -2, 1, 0, 0, 1, -2, 0, 1, 0, 0, 0)^t.
\]
Since $B_N(\lambda)$ are commutative, the remaining solution provides the same vector.
This vector gives the representation $\mathbf{1}$.
\qed
\end{example}

Recall that we have the irreducible decomposition
$
\mathfrak{H}_4=\mathbf{5}\oplus\mathbf{3}^{\oplus 3}\oplus\mathbf{1}^{\oplus 2}$.
In particular, the Hamiltonian $\mathcal{H}_4$ has one more eigenvector
\[
(0,0,0,1,0,0,-1,0,0,-1,0,0,1,0,0,0)^t
\]
or, equivalently,
\begin{align*}
|++--\rangle-|+--+\rangle
-|-++-\rangle+|--++\rangle.
\end{align*}
The missing eigenvector, which should correspond to the solutions
$\{i/2,-i/2\}$ and $\{-i/2,i/2\}$, can be found by a ``refinement"
of the Bethe ansatz method.
Below we follow Nepomechie--Wang's arguments
\cite{NepomechieWang2013}\footnote{In the case of $\ell=2$,
their result coincides with the result of \cite{EKS:1992}, equation (26).}.

We call the following type of solutions {\bf singular}:
\begin{align}
\label{eq:general_singular_solution}
\left\{
\frac{i}{2},-\frac{i}{2},\lambda_3,\ldots,\lambda_\ell
\right\}.
\end{align}
Recall that if $\{\lambda_1,\ldots,\lambda_\ell\}$ is a solution to
the Bethe ansatz equations, we have
\begin{align}
\mathcal{H}_N\Psi_N(\lambda_1,\ldots,\lambda_\ell)=
\mathcal{E}_{\lambda_1,\ldots,\lambda_\ell}
\Psi_N(\lambda_1,\ldots,\lambda_\ell),\qquad
\mathcal{E}_{\lambda_1,\ldots,\lambda_\ell}:=-\frac{J}{2}\sum_{j=1}^\ell\frac{1}{\lambda_j^2+\frac{1}{4}}.
\end{align}
Therefore singular solutions correspond to divergent energy eigenvalues.
For singular solutions, let us consider the following regularization
\begin{align}
\label{eq:regularization_of_Nepomechie}
\lambda_1=\frac{i}{2}+\epsilon+c\,\epsilon^N,\qquad
\lambda_2=-\frac{i}{2}+\epsilon.
\end{align}
We remark that the same regularization method was also noted in \cite{Beisert}, equation (3.4).
In the case of $N=4$, we obtain the following result
\begin{align*}
\lim_{\epsilon\rightarrow 0}\frac{1}{\epsilon^4}
B_4\!\left(\frac{i}{2}+\epsilon+c\,\epsilon^4\right)B_4\!\left(-\frac{i}{2}+\epsilon\right)|0\rangle_4
=(0,0,0,2,0,0,-2,0,0,ic,0,0,2,0,0,0)^t.
\end{align*}
If $c=2i$, we obtain the correct eigenvector.

In general, Nepomechie--Wang's prescription is as follows.
We start from the general singular solution (\ref{eq:general_singular_solution}).
If a singular solution satisfies the relation
\begin{align}
\left(
-\prod_{j=3}^\ell
\frac{\lambda_j+\frac{i}{2}}{\lambda_j-\frac{i}{2}}
\right)^N=1,
\end{align}
we call {\bf physical singular solution}.
We define the number $c$ as follows
\begin{align}
c=2i^{N+1}\prod_{j=3}^\ell
\frac{\lambda_j+\frac{3i}{2}}{\lambda_j-\frac{i}{2}}.
\end{align}
According to \cite{NepomechieWang2013},
\begin{align}
\lim_{\epsilon\rightarrow 0}\frac{1}{\epsilon^N}
B_N\!\left(\frac{i}{2}+\epsilon+c\,\epsilon^N\right)B_N\!\left(-\frac{i}{2}+\epsilon\right)
B_N(\lambda_3)\cdots B_N(\lambda_\ell)
|0\rangle_N
\end{align}
gives a non-zero eigenvector of the Hamiltonian.

Then the main conjecture of \cite{NepomechieWang2013} is as follows.
\begin{conjecture}
Regular solutions and physical singular solutions of the Bethe ansatz equations
provide all the highest weight vectors of $\mathfrak{H}_N$.
\qed
\end{conjecture}
In \cite{HaoNepomechieSommese},
this conjecture is verified up to $N=14$ by an extensive numerical computation.

\section{Physical singular solutions to the Bethe ansatz equations and rigged configurations}
\label{sec:RC}
\subsection{Rigged Configurations---${\mathfrak{sl}}_2$-case}

Rigged configurations have been introduced by the first author and N. Reshetikhin
in the beginning of 80's of the last century as a consequence of application
of the {\it String Conjecture} to the problem  of counting the number of
``{\it physically interesting solutions}''  to the Bethe ansatz equations, \cite{Kir}.
In this Section we remind to the reader a definition and some basic result in the case of 
${\mathfrak{sl}}_2$ Heisenberg model.

In a few words, a rigged configuration (in the case of ${\mathfrak{sl}}_2$) 
is a pair $(\nu, J)$, where $\nu$ is a partition, and $J$ is a weakly 
increasing sequence of non-negative integer numbers of the length equals to 
the number of parts of a partition $\nu$. A pair $(\nu, J)$ has to satisfy
certain conditions depending on a type of the Heisenberg model we are 
interested in. The starting data is a collection of positive integers (spins)
${\bf \mu} = (\mu_1,\ldots,\mu_N)$ and a partition $\eta= (\eta_1,
\eta_2)$ $(\eta_1,\eta_2 \ge 0)$ such that $\eta_1+\eta_2=\sum_{j=1}^r\mu_j$.
In the language of the Heisenberg model, $\mu$ specifies the shape
of the tensor product of the space of states and $\eta$
specifies the type of the Bethe vectors.
For example, if we consider a Bethe vector $B_N(\lambda_1)\cdots B_N(\lambda_\ell)|0\rangle_N$
of the tensor product of spin-$\frac{1}{2}$ representations
$\mathfrak{H}_N=(\mathbb{C}^2)^{\otimes N}$, we have
$\mu=(\overbrace{1,1,\cdots ,1}^N)=(1^N)$ and
$\eta=(\eta_1,\eta_2)=(N-\ell,\ell)$.

\begin{definition} Given $\mu$ and $\eta$ as above, a configuration of 
type $\eta$ is a partition $\nu=(\nu_1,\ldots,\nu_s)$ such that 
$\sum_{j=1}^s \nu_j =\eta_2$.
\qed
\end{definition}

\begin{definition}
For a given configuration of type $\mu=(\mu_{1}, \ldots, \mu_N)$
and $\nu=(\nu_1,\ldots,\nu_s)$, define the so-called 
{\bf vacancy numbers} $P_k(\nu)$ as follows
\begin{align}
\label{eq:def:vacancy_numbers}
P_{k}(\nu)= \sum_{j=1}^{N}\min(k,\mu_j) -2 \sum_{j=1}^{s}\min(k,\nu_j),\qquad
k\in\mathbb{Z}_{\geq 0}.
\end{align}
A configuration $\nu$ of type $(\mu,\eta)$ is called
{\bf admissible}, if all vacancy numbers $\{ P_k(\nu) \}$ are non-negative.
\qed
\end{definition}

In the following, we will freely identify the partitions and the Young diagrams.
Then the number $\sum_{j=1}^{s}\min(k,\nu_j)$ is the number of boxes within
the first $k$ columns of $\nu$.
Here we prepare useful notation concerning the Young diagrams.
Let $m_k(\nu)$ be the number of length $k$ rows of $\nu$,
that is, $m_k(\nu)=m_k= \#\{ j\,|\,\nu_j=k \}$ and
let $|\nu|$ be the total number of boxes of the corresponding
Young diagram; $|\nu|=\sum_{j=1}^{s}\nu_j$.

We remark that in the case of the length $N$ spin-$\frac{1}{2}$ model,
the definition of the vacancy numbers
(\ref{eq:def:vacancy_numbers}) is simply
\begin{align}
P_{k}(\nu)= N-2 \sum_{j=1}^{s}\min(k,\nu_j),\qquad
k\in\mathbb{Z}_{> 0}.
\end{align}
In particular, it should be reminded that $P_k(\nu)$ depends on the data $N$.
In general, the vacancy numbers depend on the data $\mu$,
although the abbreviated symbol $P_{k}(\nu)$ is commonly used in
the rigged configurations literature.

\begin{example}
If $\mu=(\overbrace{1,1,\cdots ,1}^N)$,
every partition $\nu$ such that $|\nu|$ does not exceed $N/2$ is admissible.
On the other hand, if $\mu=(2,2)$, $\nu=\emptyset, (1), (2)$ are admissible
but $\nu=(1,1)$ is not admissible since we have $P_1((1,1))=-2$.
\end{example}

\begin{remark}
It seems appropriate to address the following supplementary information,
though the relation with the present problem is yet unclear.
There is a canonical bijection between the set of rigged configurations
and tensor products of crystals (see \cite{S:web} for a Mathematica implementation).
Then the rigged configurations defined here correspond to the highest weight
element of crystals.
In order to include non-highest weight crystals, we need to relax the admissibility
of the configurations.
See, for example, \cite{S:2014} and references therein for details.
\qed
\end{remark}

\begin{definition} A rigged configuration $(\nu,J)$ of type $(\mu, \eta)$ is an
admissible configuration $\nu$ of type $\eta$ together with a weakly increasing 
sequence of non-negative integers 
\begin{align}
0 \le J_{k,1} \le J_{k,2} \le\cdots\le J_{k,m_{k}} \le P_k(\nu),\qquad
k=1,2, \ldots .
\end{align}
We call the integers $J_{k,\alpha}$ $(\alpha=1,\ldots,m_k)$ the {\bf riggings}
associated to the length $k$ rows of the partition $\nu$.
We denote by $\mathrm{RC}(\mu, \nu)$ the set of rigged configurations
with specific $\mu$ and $\nu$.
\qed
\end{definition}
\begin{definition}  Define the flip map 
\begin{align}
\kappa: \mathrm{RC}(\mu,\nu) \longrightarrow \mathrm{RC}(\mu,\nu)
\end{align}
as follows: $\kappa(\nu)=\nu$, and 
\begin{align}
\kappa(J_{k,\alpha}) = P_k(\nu)-J_{k, m_k -\alpha+1}
\end{align}
for all $k\in\mathbb{Z}_{>0}$ and $\alpha =1,\ldots,m_k$.
\qed
\end{definition}

It is convenient to regard the rigged configuration $(\nu,J)$
as a collection of data $(k,J_{k,\alpha})$
($k\in\mathbb{Z}_{>0}$, $\alpha =1,\ldots,m_k(\nu)$) which we call strings.
Then the flip map is the map
\[
(k,J_{k,\alpha})\longmapsto (k,P_k(\nu)-J_{k,\alpha})
\]
together with certain reordering of the strings to make the riggings satisfy
the weakly increasing condition.
However, we note that the order of the strings is not essential in the rigged configuration theory.

\begin{example}
For $\mu=(1^{16})$, the following two rigged configurations are related under the flip map $\kappa$.
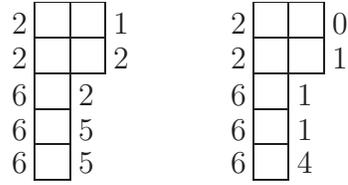
\begin{center}
\unitlength 12pt
\begin{picture}(1.5,5.5)(-0.7,0)
\put(-0.7,4.6){$2$}
\put(-0.7,3.5){$2$}
\put(-0.7,2.35){$6$}
\put(-0.7,1.25){$6$}
\put(-0.7,0.2){$6$}
\put(0,0){\yng(2,2,1,1,1)}
\put(2.5,4.6){$1$}
\put(2.5,3.5){$2$}
\put(1.4,2.35){$2$}
\put(1.4,1.25){$5$}
\put(1.4,0.2){$5$}
\end{picture}
\hspace{20mm}
\begin{picture}(1.5,5.5)(-0.7,0)
\put(-0.7,4.6){$2$}
\put(-0.7,3.5){$2$}
\put(-0.7,2.35){$6$}
\put(-0.7,1.25){$6$}
\put(-0.7,0.2){$6$}
\put(0,0){\yng(2,2,1,1,1)}
\put(2.5,4.6){$0$}
\put(2.5,3.5){$1$}
\put(1.4,2.35){$1$}
\put(1.4,1.25){$1$}
\put(1.4,0.2){$4$}
\end{picture}
\end{center}
Here we depict the configuration $\nu$ by the Young diagram.
For the string $(k,J_{k,\alpha})$, we put $P_k(\nu)$ (resp. $J_{k,\alpha}$)
on the left (resp. right) of the corresponding length $k$ row of the diagram.
\qed
\end{example}

Our fundamental observation is that the rigged configurations
provide a nice parameterization of the combination of both regular solutions
and physical singular solutions to the Bethe ansatz equations.
As we explain in the following examples,
the basic idea is to identify the string $(k,J_{k,\alpha})$ of the rigged configuration
with the collection of $k$ solutions which have (almost) same real part
specified by $J_{k,\alpha}$.
We call such collection of roots as length $k$ string of solutions.
Let us tentatively suppose that the larger rigging corresponds
to the rightwards string of solutions.
However there is an ambiguity described in Conjecture \ref{conj:main} ({\bf A}) below.

\begin{example}
Consider the spin-$\frac{1}{2}$ case of $N=6$ and $\ell=3$.
Then we have the following five solutions to the Bethe ansatz equations (\ref{eq:Bethe_ansatz})
which are depicted on the complex plain.
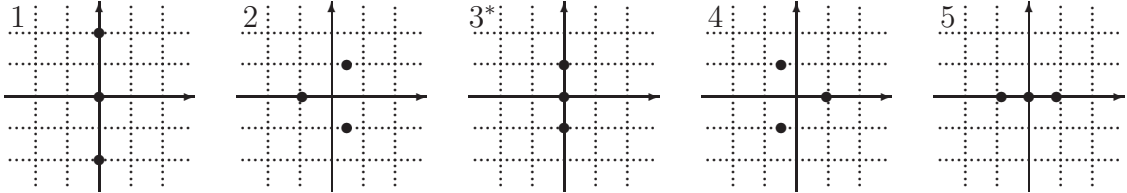
\begin{center}
\unitlength 12pt
\begin{picture}(7,6)
\put(0.2,5.2){$1$}
\put(0,3){\vector(1,0){6}}
\put(3,0){\vector(0,1){6}}
\put(3,1){\circle*{0.3}}
\put(3,3){\circle*{0.3}}
\put(3,5){\circle*{0.3}}
\multiput(1,0.21)(0,0.2){29}{\circle*{0.07}}
\multiput(2,0.21)(0,0.2){29}{\circle*{0.07}}
\multiput(4,0.21)(0,0.2){29}{\circle*{0.07}}
\multiput(5,0.21)(0,0.2){29}{\circle*{0.07}}
\multiput(0.2,1.01)(0.2,0){29}{\circle*{0.07}}
\multiput(0.2,2.01)(0.2,0){29}{\circle*{0.07}}
\multiput(0.2,4.01)(0.2,0){29}{\circle*{0.07}}
\multiput(0.2,5.01)(0.2,0){29}{\circle*{0.07}}
\end{picture}
\begin{picture}(7,6)
\put(0.2,5.2){$2$}
\put(0,3){\vector(1,0){6}}
\put(3,0){\vector(0,1){6}}
\put(2.06,3){\circle*{0.3}}
\put(3.48,2){\circle*{0.3}}
\put(3.48,4){\circle*{0.3}}
\multiput(1,0.21)(0,0.2){29}{\circle*{0.07}}
\multiput(2,0.21)(0,0.2){29}{\circle*{0.07}}
\multiput(4,0.21)(0,0.2){29}{\circle*{0.07}}
\multiput(5,0.21)(0,0.2){29}{\circle*{0.07}}
\multiput(0.2,1.01)(0.2,0){29}{\circle*{0.07}}
\multiput(0.2,2.01)(0.2,0){29}{\circle*{0.07}}
\multiput(0.2,4.01)(0.2,0){29}{\circle*{0.07}}
\multiput(0.2,5.01)(0.2,0){29}{\circle*{0.07}}
\end{picture}
\begin{picture}(7,6)
\put(0,5.2){$3^\ast$}
\put(0,3){\vector(1,0){6}}
\put(3,0){\vector(0,1){6}}
\put(3,3){\circle*{0.3}}
\put(3,2){\circle*{0.3}}
\put(3,4){\circle*{0.3}}
\multiput(1,0.21)(0,0.2){29}{\circle*{0.07}}
\multiput(2,0.21)(0,0.2){29}{\circle*{0.07}}
\multiput(4,0.21)(0,0.2){29}{\circle*{0.07}}
\multiput(5,0.21)(0,0.2){29}{\circle*{0.07}}
\multiput(0.2,1.01)(0.2,0){29}{\circle*{0.07}}
\multiput(0.2,2.01)(0.2,0){29}{\circle*{0.07}}
\multiput(0.2,4.01)(0.2,0){29}{\circle*{0.07}}
\multiput(0.2,5.01)(0.2,0){29}{\circle*{0.07}}
\end{picture}
\begin{picture}(7,6)
\put(0.2,5.2){$4$}
\put(0,3){\vector(1,0){6}}
\put(3,0){\vector(0,1){6}}
\put(3.94,3){\circle*{0.3}}
\put(2.52,2){\circle*{0.3}}
\put(2.52,4){\circle*{0.3}}
\multiput(1,0.21)(0,0.2){29}{\circle*{0.07}}
\multiput(2,0.21)(0,0.2){29}{\circle*{0.07}}
\multiput(4,0.21)(0,0.2){29}{\circle*{0.07}}
\multiput(5,0.21)(0,0.2){29}{\circle*{0.07}}
\multiput(0.2,1.01)(0.2,0){29}{\circle*{0.07}}
\multiput(0.2,2.01)(0.2,0){29}{\circle*{0.07}}
\multiput(0.2,4.01)(0.2,0){29}{\circle*{0.07}}
\multiput(0.2,5.01)(0.2,0){29}{\circle*{0.07}}
\end{picture}
\begin{picture}(7,6)
\put(0.2,5.2){$5$}
\put(0,3){\vector(1,0){6}}
\put(3,0){\vector(0,1){6}}
\put(3,3){\circle*{0.3}}
\put(3.86,3){\circle*{0.3}}
\put(2.14,3){\circle*{0.3}}
\multiput(1,0.21)(0,0.2){29}{\circle*{0.07}}
\multiput(2,0.21)(0,0.2){29}{\circle*{0.07}}
\multiput(4,0.21)(0,0.2){29}{\circle*{0.07}}
\multiput(5,0.21)(0,0.2){29}{\circle*{0.07}}
\multiput(0.2,1.01)(0.2,0){29}{\circle*{0.07}}
\multiput(0.2,2.01)(0.2,0){29}{\circle*{0.07}}
\multiput(0.2,4.01)(0.2,0){29}{\circle*{0.07}}
\multiput(0.2,5.01)(0.2,0){29}{\circle*{0.07}}
\end{picture}
\end{center}
Here the spacing of the dotted lines is $0.5$ and the label with asterisk
($3^\ast$ in this case) means that the solution is singular and physical.
These solutions correspond to the rigged configurations as in the following table.
\begin{center}
\begin{tabular}{|lll|}
\hline
label&values of the roots&rigged configurations\\
\hline
$1$&$0,\pm i$&
{\unitlength 12pt
\begin{picture}(3,1.5)(-0.7,0)
\put(-0.7,0.2){$0$}
\put(0,0){\yng(3)}
\put(3.7,0.2){$0$}
\end{picture}}\\
$2$&$-0.47,0.24\pm 0.5i$&
{\unitlength 12pt
\begin{picture}(2.5,2.5)(-0.7,0)
\put(-0.7,1.2){$0$}
\put(-0.7,0.2){$2$}
\put(0,0){\yng(2,1)}
\put(2.6,1.2){$0$}
\put(1.4,0.2){$0$}
\end{picture}}\\
$3^\ast$&$0,\pm 0.5i$&
{\unitlength 12pt
\begin{picture}(2.5,2.5)(-0.7,0)
\put(-0.7,1.2){$0$}
\put(-0.7,0.2){$2$}
\put(0,0){\yng(2,1)}
\put(2.6,1.2){$0$}
\put(1.4,0.2){$1$}
\end{picture}}\\
$4$&$0.47,-0.24\pm 0.5i$&
{\unitlength 12pt
\begin{picture}(2.5,2.5)(-0.7,0)
\put(-0.7,1.2){$0$}
\put(-0.7,0.2){$2$}
\put(0,0){\yng(2,1)}
\put(2.6,1.2){$0$}
\put(1.4,0.2){$2$}
\end{picture}}\\
$5$&$0,\pm 0.43$&
{\unitlength 12pt
\begin{picture}(1.5,3.5)(-0.7,0)
\put(-0.7,2.3){$0$}
\put(-0.7,1.2){$0$}
\put(-0.7,0.2){$0$}
\put(0,0){\yng(1,1,1)}
\put(1.4,2.3){$0$}
\put(1.4,1.2){$0$}
\put(1.4,0.2){$0$}
\end{picture}}\\\hline
\end{tabular}
\end{center}

Let us explain the above correspondence in more detail.
The condition $N=6$ and $\ell=3$ means that $\mu=(1^6)$ and $|\nu|=3$.
Then the admissible configurations are $\nu=(3), (2,1), (1^3)$.
Since $P_3((3))=P_1((1^3))=0$, the corresponding riggings are 0.
Therefore, for the cases $\nu=(3)$ and $\nu=(1^3)$,
there is no ambiguity in the correspondence between the rigged configurations
and the solutions to the Bethe ansatz equations.

Let us consider the case $\nu=(2,1)$.
Since $P_2((2,1))=0$, there is no choice for the length 2 string of solutions.
On the other hand, since $P_1((2,1))=2$, we have three possibilities for
the rigging of the length 1 row of $\nu$.
If we make the assumption such that the larger rigging corresponds to
the rightwards string of solutions, we obtain the above correspondence.
Notably, the above correspondence is compatible with the real parts
of the length 2 string of solutions.
\qed
\end{example}

\begin{example}
Consider the spin-$\frac{1}{2}$ case of $N=8$ and $\ell=4$ \cite{HaoNepomechieSommese}.
We keep the notations of the previous example.
The following solutions corresponds to rigged configurations
with uniquely determined riggings.
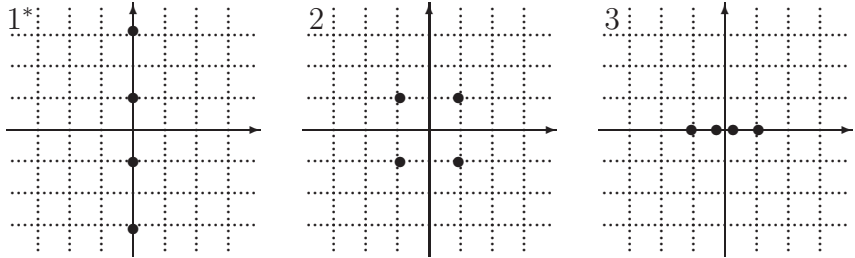
\begin{center}
\unitlength 12pt
\begin{picture}(9,8)
\put(-0.0,7.2){$1^\ast$}
\put(0,4){\vector(1,0){8}}
\put(4,0){\vector(0,1){8}}
\put(4,3){\circle*{0.3}}
\put(4,5){\circle*{0.3}}
\put(4,7.12){\circle*{0.3}}
\put(4,0.88){\circle*{0.3}}
\multiput(1,0.21)(0,0.2){39}{\circle*{0.07}}
\multiput(2,0.21)(0,0.2){39}{\circle*{0.07}}
\multiput(3,0.21)(0,0.2){39}{\circle*{0.07}}
\multiput(5,0.21)(0,0.2){39}{\circle*{0.07}}
\multiput(6,0.21)(0,0.2){39}{\circle*{0.07}}
\multiput(7,0.21)(0,0.2){39}{\circle*{0.07}}
\multiput(0.2,1.01)(0.2,0){39}{\circle*{0.07}}
\multiput(0.2,2.01)(0.2,0){39}{\circle*{0.07}}
\multiput(0.2,3.01)(0.2,0){39}{\circle*{0.07}}
\multiput(0.2,5.01)(0.2,0){39}{\circle*{0.07}}
\multiput(0.2,6.01)(0.2,0){39}{\circle*{0.07}}
\multiput(0.2,7.01)(0.2,0){39}{\circle*{0.07}}
\end{picture}
\begin{picture}(9,8)
\put(0.2,7.2){$2$}
\put(0,4){\vector(1,0){8}}
\put(4,0){\vector(0,1){8}}
\put(3.08,3){\circle*{0.3}}
\put(3.08,5){\circle*{0.3}}
\put(4.92,3){\circle*{0.3}}
\put(4.92,5){\circle*{0.3}}
\multiput(1,0.21)(0,0.2){39}{\circle*{0.07}}
\multiput(2,0.21)(0,0.2){39}{\circle*{0.07}}
\multiput(3,0.21)(0,0.2){39}{\circle*{0.07}}
\multiput(5,0.21)(0,0.2){39}{\circle*{0.07}}
\multiput(6,0.21)(0,0.2){39}{\circle*{0.07}}
\multiput(7,0.21)(0,0.2){39}{\circle*{0.07}}
\multiput(0.2,1.01)(0.2,0){39}{\circle*{0.07}}
\multiput(0.2,2.01)(0.2,0){39}{\circle*{0.07}}
\multiput(0.2,3.01)(0.2,0){39}{\circle*{0.07}}
\multiput(0.2,5.01)(0.2,0){39}{\circle*{0.07}}
\multiput(0.2,6.01)(0.2,0){39}{\circle*{0.07}}
\multiput(0.2,7.01)(0.2,0){39}{\circle*{0.07}}
\end{picture}
\begin{picture}(9,8)
\put(0.2,7.2){$3$}
\put(0,4){\vector(1,0){8}}
\put(4,0){\vector(0,1){8}}
\put(4.26,4){\circle*{0.3}}
\put(5.06,4){\circle*{0.3}}
\put(3.74,4){\circle*{0.3}}
\put(2.94,4){\circle*{0.3}}
\multiput(1,0.21)(0,0.2){39}{\circle*{0.07}}
\multiput(2,0.21)(0,0.2){39}{\circle*{0.07}}
\multiput(3,0.21)(0,0.2){39}{\circle*{0.07}}
\multiput(5,0.21)(0,0.2){39}{\circle*{0.07}}
\multiput(6,0.21)(0,0.2){39}{\circle*{0.07}}
\multiput(7,0.21)(0,0.2){39}{\circle*{0.07}}
\multiput(0.2,1.01)(0.2,0){39}{\circle*{0.07}}
\multiput(0.2,2.01)(0.2,0){39}{\circle*{0.07}}
\multiput(0.2,3.01)(0.2,0){39}{\circle*{0.07}}
\multiput(0.2,5.01)(0.2,0){39}{\circle*{0.07}}
\multiput(0.2,6.01)(0.2,0){39}{\circle*{0.07}}
\multiput(0.2,7.01)(0.2,0){39}{\circle*{0.07}}
\end{picture}
\end{center}
These solutions correspond to the following rigged configurations.
\begin{center}
\begin{tabular}{|lll|}
\hline
label&values of the roots&rigged configuration\\
\hline
$1^\ast$&$\pm 0.5i,\pm1.56i$&
{\unitlength 12pt
\begin{picture}(5,1.5)(-0.7,0)
\put(-0.7,0.2){$0$}
\put(0,0){\yng(4)}
\put(4.8,0.2){$0$}
\end{picture}}\\
$2$&$\pm 0.46\pm 0.5i$&
{\unitlength 12pt
\begin{picture}(5,2.5)(-0.7,0)
\put(-0.7,1.2){$0$}
\put(-0.7,0.2){$0$}
\put(0,0){\yng(2,2)}
\put(2.5,1.2){$0$}
\put(2.5,0.2){$0$}
\end{picture}}\\
$3$&$\pm 0.13,\pm 0.53$&
{\unitlength 12pt
\begin{picture}(5,4.5)(-0.7,0)
\put(-0.7,3.5){$0$}
\put(-0.7,2.4){$0$}
\put(-0.7,1.3){$0$}
\put(-0.7,0.2){$0$}
\put(0,0){\yng(1,1,1,1)}
\put(1.4,3.5){$0$}
\put(1.4,2.4){$0$}
\put(1.4,1.3){$0$}
\put(1.4,0.2){$0$}
\end{picture}}\\
\hline
\end{tabular}
\end{center}

Consider the following five solutions.
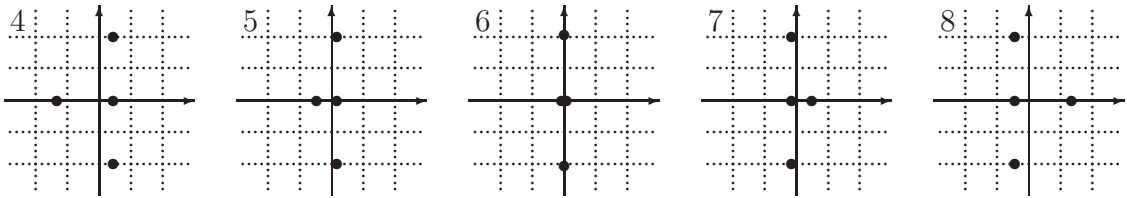
\begin{center}
\unitlength 12pt
\begin{picture}(7,6)
\put(0.2,5.2){$4$}
\put(0,3){\vector(1,0){6}}
\put(3,0){\vector(0,1){6}}
\put(3.44,1){\circle*{0.3}}
\put(3.44,3){\circle*{0.3}}
\put(3.44,5){\circle*{0.3}}
\put(1.66,3){\circle*{0.3}}
\multiput(1,0.21)(0,0.2){29}{\circle*{0.07}}
\multiput(2,0.21)(0,0.2){29}{\circle*{0.07}}
\multiput(4,0.21)(0,0.2){29}{\circle*{0.07}}
\multiput(5,0.21)(0,0.2){29}{\circle*{0.07}}
\multiput(0.2,1.01)(0.2,0){29}{\circle*{0.07}}
\multiput(0.2,2.01)(0.2,0){29}{\circle*{0.07}}
\multiput(0.2,4.01)(0.2,0){29}{\circle*{0.07}}
\multiput(0.2,5.01)(0.2,0){29}{\circle*{0.07}}
\end{picture}
\begin{picture}(7,6)
\put(0.2,5.2){$5$}
\put(0,3){\vector(1,0){6}}
\put(3,0){\vector(0,1){6}}
\put(3.16,1){\circle*{0.3}}
\put(3.16,3){\circle*{0.3}}
\put(3.16,5){\circle*{0.3}}
\put(2.52,3){\circle*{0.3}}
\multiput(1,0.21)(0,0.2){29}{\circle*{0.07}}
\multiput(2,0.21)(0,0.2){29}{\circle*{0.07}}
\multiput(4,0.21)(0,0.2){29}{\circle*{0.07}}
\multiput(5,0.21)(0,0.2){29}{\circle*{0.07}}
\multiput(0.2,1.01)(0.2,0){29}{\circle*{0.07}}
\multiput(0.2,2.01)(0.2,0){29}{\circle*{0.07}}
\multiput(0.2,4.01)(0.2,0){29}{\circle*{0.07}}
\multiput(0.2,5.01)(0.2,0){29}{\circle*{0.07}}
\end{picture}
\begin{picture}(7,6)
\put(0.2,5.2){$6$}
\put(0,3){\vector(1,0){6}}
\put(3,0){\vector(0,1){6}}
\put(2.92,3){\circle*{0.3}}
\put(3.08,3){\circle*{0.3}}
\put(3,5.06){\circle*{0.3}}
\put(3,0.94){\circle*{0.3}}
\multiput(1,0.21)(0,0.2){29}{\circle*{0.07}}
\multiput(2,0.21)(0,0.2){29}{\circle*{0.07}}
\multiput(4,0.21)(0,0.2){29}{\circle*{0.07}}
\multiput(5,0.21)(0,0.2){29}{\circle*{0.07}}
\multiput(0.2,1.01)(0.2,0){29}{\circle*{0.07}}
\multiput(0.2,2.01)(0.2,0){29}{\circle*{0.07}}
\multiput(0.2,4.01)(0.2,0){29}{\circle*{0.07}}
\multiput(0.2,5.01)(0.2,0){29}{\circle*{0.07}}
\end{picture}
\begin{picture}(7,6)
\put(0.2,5.2){$7$}
\put(0,3){\vector(1,0){6}}
\put(3,0){\vector(0,1){6}}
\put(3.48,3){\circle*{0.3}}
\put(2.84,1){\circle*{0.3}}
\put(2.84,3){\circle*{0.3}}
\put(2.84,5){\circle*{0.3}}
\multiput(1,0.21)(0,0.2){29}{\circle*{0.07}}
\multiput(2,0.21)(0,0.2){29}{\circle*{0.07}}
\multiput(4,0.21)(0,0.2){29}{\circle*{0.07}}
\multiput(5,0.21)(0,0.2){29}{\circle*{0.07}}
\multiput(0.2,1.01)(0.2,0){29}{\circle*{0.07}}
\multiput(0.2,2.01)(0.2,0){29}{\circle*{0.07}}
\multiput(0.2,4.01)(0.2,0){29}{\circle*{0.07}}
\multiput(0.2,5.01)(0.2,0){29}{\circle*{0.07}}
\end{picture}
\begin{picture}(7,6)
\put(0.2,5.2){$8$}
\put(0,3){\vector(1,0){6}}
\put(3,0){\vector(0,1){6}}
\put(2.56,1){\circle*{0.3}}
\put(2.56,3){\circle*{0.3}}
\put(2.56,5){\circle*{0.3}}
\put(4.34,3){\circle*{0.3}}
\multiput(1,0.21)(0,0.2){29}{\circle*{0.07}}
\multiput(2,0.21)(0,0.2){29}{\circle*{0.07}}
\multiput(4,0.21)(0,0.2){29}{\circle*{0.07}}
\multiput(5,0.21)(0,0.2){29}{\circle*{0.07}}
\multiput(0.2,1.01)(0.2,0){29}{\circle*{0.07}}
\multiput(0.2,2.01)(0.2,0){29}{\circle*{0.07}}
\multiput(0.2,4.01)(0.2,0){29}{\circle*{0.07}}
\multiput(0.2,5.01)(0.2,0){29}{\circle*{0.07}}
\end{picture}
\end{center}
These solutions correspond to the following rigged configurations.
\begin{center}
\unitlength 12pt
\begin{picture}(5,2.5)(-0.7,0)
\put(-0.7,1.2){$0$}
\put(-0.7,0.2){$4$}
\put(0,0){\yng(3,1)}
\put(3.7,1.2){$0$}
\put(1.4,0.2){$r$}
\end{picture}
\end{center}
\begin{center}
\begin{tabular}{|llc|}
\hline
label&values of the roots&value of $r$\\
\hline
4&$-0.67,0.22,0.22\pm i$&0\\
5&$-0.24,0.08\pm 1.01i,0.08$&1\\
6&$\pm 0.04,\pm 1.03i$&2\\
7&$-0.08,-0.08\pm 1.01i,0.24$&3\\
8&$-0.22\pm i,-0.22,0.67$&4\\
\hline
\end{tabular}
\end{center}

Finally let us consider the following six solutions.

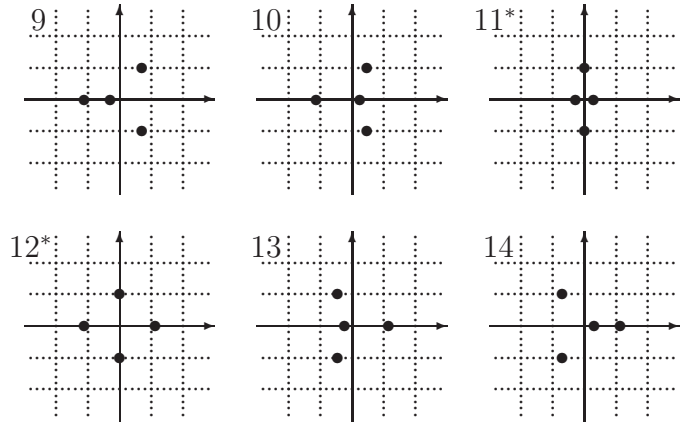
\begin{center}
\unitlength 12pt
\begin{picture}(7,6)
\put(0.2,5.2){$9$}
\put(0,3){\vector(1,0){6}}
\put(3,0){\vector(0,1){6}}
\put(3.7,2){\circle*{0.3}}
\put(3.7,4){\circle*{0.3}}
\put(2.7,3){\circle*{0.3}}
\put(1.88,3){\circle*{0.3}}
\multiput(1,0.21)(0,0.2){29}{\circle*{0.07}}
\multiput(2,0.21)(0,0.2){29}{\circle*{0.07}}
\multiput(4,0.21)(0,0.2){29}{\circle*{0.07}}
\multiput(5,0.21)(0,0.2){29}{\circle*{0.07}}
\multiput(0.2,1.01)(0.2,0){29}{\circle*{0.07}}
\multiput(0.2,2.01)(0.2,0){29}{\circle*{0.07}}
\multiput(0.2,4.01)(0.2,0){29}{\circle*{0.07}}
\multiput(0.2,5.01)(0.2,0){29}{\circle*{0.07}}
\end{picture}
\begin{picture}(7,6)
\put(-0.2,5.2){$10$}
\put(0,3){\vector(1,0){6}}
\put(3,0){\vector(0,1){6}}
\put(3.46,4){\circle*{0.3}}
\put(3.46,2){\circle*{0.3}}
\put(3.24,3){\circle*{0.3}}
\put(1.86,3){\circle*{0.3}}
\multiput(1,0.21)(0,0.2){29}{\circle*{0.07}}
\multiput(2,0.21)(0,0.2){29}{\circle*{0.07}}
\multiput(4,0.21)(0,0.2){29}{\circle*{0.07}}
\multiput(5,0.21)(0,0.2){29}{\circle*{0.07}}
\multiput(0.2,1.01)(0.2,0){29}{\circle*{0.07}}
\multiput(0.2,2.01)(0.2,0){29}{\circle*{0.07}}
\multiput(0.2,4.01)(0.2,0){29}{\circle*{0.07}}
\multiput(0.2,5.01)(0.2,0){29}{\circle*{0.07}}
\end{picture}
\begin{picture}(7,6)
\put(-0.5,5.2){$11^\ast$}
\put(0,3){\vector(1,0){6}}
\put(3,0){\vector(0,1){6}}
\put(3,2){\circle*{0.3}}
\put(3,4){\circle*{0.3}}
\put(3.28,3){\circle*{0.3}}
\put(2.72,3){\circle*{0.3}}
\multiput(1,0.21)(0,0.2){29}{\circle*{0.07}}
\multiput(2,0.21)(0,0.2){29}{\circle*{0.07}}
\multiput(4,0.21)(0,0.2){29}{\circle*{0.07}}
\multiput(5,0.21)(0,0.2){29}{\circle*{0.07}}
\multiput(0.2,1.01)(0.2,0){29}{\circle*{0.07}}
\multiput(0.2,2.01)(0.2,0){29}{\circle*{0.07}}
\multiput(0.2,4.01)(0.2,0){29}{\circle*{0.07}}
\multiput(0.2,5.01)(0.2,0){29}{\circle*{0.07}}
\end{picture}
\end{center}

\begin{center}
\unitlength 12pt
\begin{picture}(7,6)
\put(-0.5,5.2){$12^\ast$}
\put(0,3){\vector(1,0){6}}
\put(3,0){\vector(0,1){6}}
\put(3,4){\circle*{0.3}}
\put(3,2){\circle*{0.3}}
\put(4.12,3){\circle*{0.3}}
\put(1.88,3){\circle*{0.3}}
\multiput(1,0.21)(0,0.2){29}{\circle*{0.07}}
\multiput(2,0.21)(0,0.2){29}{\circle*{0.07}}
\multiput(4,0.21)(0,0.2){29}{\circle*{0.07}}
\multiput(5,0.21)(0,0.2){29}{\circle*{0.07}}
\multiput(0.2,1.01)(0.2,0){29}{\circle*{0.07}}
\multiput(0.2,2.01)(0.2,0){29}{\circle*{0.07}}
\multiput(0.2,4.01)(0.2,0){29}{\circle*{0.07}}
\multiput(0.2,5.01)(0.2,0){29}{\circle*{0.07}}
\end{picture}
\begin{picture}(7,6)
\put(-0.2,5.2){$13$}
\put(0,3){\vector(1,0){6}}
\put(3,0){\vector(0,1){6}}
\put(2.54,2){\circle*{0.3}}
\put(2.54,4){\circle*{0.3}}
\put(4.14,3){\circle*{0.3}}
\put(2.76,3){\circle*{0.3}}
\multiput(1,0.21)(0,0.2){29}{\circle*{0.07}}
\multiput(2,0.21)(0,0.2){29}{\circle*{0.07}}
\multiput(4,0.21)(0,0.2){29}{\circle*{0.07}}
\multiput(5,0.21)(0,0.2){29}{\circle*{0.07}}
\multiput(0.2,1.01)(0.2,0){29}{\circle*{0.07}}
\multiput(0.2,2.01)(0.2,0){29}{\circle*{0.07}}
\multiput(0.2,4.01)(0.2,0){29}{\circle*{0.07}}
\multiput(0.2,5.01)(0.2,0){29}{\circle*{0.07}}
\end{picture}
\begin{picture}(7,6)
\put(-0.2,5.2){$14$}
\put(0,3){\vector(1,0){6}}
\put(3,0){\vector(0,1){6}}
\put(4.12,3){\circle*{0.3}}
\put(3.3,3){\circle*{0.3}}
\put(2.3,2.0){\circle*{0.3}}
\put(2.3,4.0){\circle*{0.3}}
\multiput(1,0.21)(0,0.2){29}{\circle*{0.07}}
\multiput(2,0.21)(0,0.2){29}{\circle*{0.07}}
\multiput(4,0.21)(0,0.2){29}{\circle*{0.07}}
\multiput(5,0.21)(0,0.2){29}{\circle*{0.07}}
\multiput(0.2,1.01)(0.2,0){29}{\circle*{0.07}}
\multiput(0.2,2.01)(0.2,0){29}{\circle*{0.07}}
\multiput(0.2,4.01)(0.2,0){29}{\circle*{0.07}}
\multiput(0.2,5.01)(0.2,0){29}{\circle*{0.07}}
\end{picture}
\end{center}
Here the solutions are arranged according to the real parts of the length 2 strings of solutions.
These solutions correspond to the following rigged configurations.
\begin{center}
\unitlength 12pt
\begin{picture}(3,3.5)(-0.7,0)
\put(-0.7,2.4){$0$}
\put(-0.7,1.3){$2$}
\put(-0.7,0.2){$2$}
\put(0,0){\yng(2,1,1)}
\put(2.6,2.4){$0$}
\put(1.5,1.3){$r_1$}
\put(1.5,0.2){$r_2$}
\end{picture}
\end{center}
\begin{center}
\begin{tabular}{|llc|}
\hline
label&values of the roots&value of $(r_1,r_2)$\\
\hline
9&$-0.56,-0.14,0.35\pm 0.5i$&$(0,0)$\\
10&$-0.57,0.12,0.23\pm 0.5i$&$(0,1)$\\
$11^\ast$&$\pm 0.14,\pm 0.5i$&$(1,1)$\\
$12^\ast$&$\pm 0.56,\pm 0.5i$&$(0,2)$\\
13&$-0.23\pm 0.5i,-0.12,0.57$&$(1,2)$\\
14&$-0.35\pm 0.5i,0.14,0.56$&$(2,2)$\\
\hline
\end{tabular}
\end{center}

Let us describe the above correspondence in more detail.
As described in Conjecture \ref{conj:main}, we assume that
the flip map $\kappa$ corresponds to the multiplication of $(-1)$
to every root of the Bethe ansatz equations.
Then the solutions 11 and 12 should correspond to the rigged configurations
which are invariant under $\kappa$.
Therefore these solutions correspond to $(r_1,r_2)=(1,1)$ or $(0,2)$.
Recall that we assume that the riggings specify the positions of the corresponding
string of solutions.
Then the solution corresponding to $(r_1,r_2)=(0,2)$ must have wider spacing
between two length 1 strings of solutions compared with the one corresponding to
$(r_1,r_2)=(1,1)$.
Thus we conclude the correspondences for the solutions 11 and 12.

For the remaining solutions, if we take the map $\kappa$ into the consideration,
we need to find the correspondence for the solutions 9 and 10.
Since we assume that the riggings specify the positions of the string of solutions,
we arrive at the correspondence in the above table.
\qed
\end{example}

In Section \ref{sec:N=9&l=3} we will analyze yet another situation
involving larger values of the vacancy numbers.
Remarkably we have a natural correspondence between
the rigged configurations and strings of solutions in such general case.

\subsection{Main Conjectures}
Motivated by the examples in the previous subsection,
we propose the following conjecture.
Let $\mathrm{BA}(\ell)$ be the set of solutions $\{\lambda_1,\ldots,\lambda_\ell\}$
of the Bethe ansatz equations which are either regular or singular and physical.
\begin{conjecture}
There exist a bijection between $\mathrm{BA}(\ell)$
and the set of the rigged configurations $\mathrm{RC}(\mu,\nu)$
where the total number of the boxes of $\nu$ is $\ell$.
\qed
\end{conjecture}

Furthermore, we propose the following conjectures.

\begin{conjecture}
\label{conj:main}
${}$

\begin{enumerate}
\item[$({\bf A})$]
The map $\iota :\mathrm{BA}(\ell) \longrightarrow \mathrm{BA}(\ell)$ given by
\begin{align}
(\lambda_1,\ldots,\lambda_{\ell}) \in \mathrm{BA}(\ell) \longmapsto
(-\lambda_1,\ldots,-\lambda_{\ell})  \in \mathrm{BA}(\ell)
\end{align}
induces the flip map on the set of rigged configurations.
\end{enumerate}
{\rm In the next two conjectures we assume that the generalized Heisenberg chain 
is defined on the length $N$ tensor product of the spin $s$ representation.
In this case, we have $\mu=(\overbrace{2s,2s,\cdots ,2s}^N)$.}

\begin{enumerate}
\item[$({\bf B})$]
Assume that $N$ is even.
\begin{enumerate}
\item[$({\bf a})$]
If $2s$ is \underline{odd} and $\ell$ is even, then the set of physical singular solutions to
$\mathrm{BAE}(N,\ell)$ is in one-to-one correspondence with the
set of flip invariant rigged configurations $(\nu,J)$ such that
partition $\nu$ contains odd number of \underline{even} parts
which are greater than or equal to $2s+1$.

\item[$({\bf b})$]
If $2s$ is \underline{even} and $\ell$ is odd,
then the set of physical singular solutions to
$\mathrm{BAE}(N,\ell)$ is in one-to-one correspondence with the
set of flip invariant rigged configurations $(\nu,J)$ such that
partition $\nu$ contains odd number of \underline{odd} parts
which are greater than or equal to $2s+1$.
\end{enumerate}

\item[$({\bf C})$]
Assume that $s=\frac{1}{2}$, $N$ is even and $\ell$ is \underline{odd}.
Then the set of physical singular solutions to
$\mathrm{BAE}(N,\ell)$ is in one-to-one correspondence with the
set of flip invariant rigged configurations $(\nu,J)$ such that
partition $\nu$ contains odd number of even parts of the lengths longer than $2s+1$,
\underline{and} if the number $m_k(\nu)\geq 3$ is odd and the corresponding
vacancy number $P_k(\nu)>0$ is divisible by 4, then the rigging
\begin{align}
J_{k,1}=J_{k,2}=\cdots =J_{k,m_k}=\frac{P_k(\nu)}{2}
\end{align}
is forbidden.
\qed
\end{enumerate}
\end{conjecture}

\begin{example}${}$

\begin{itemize}
\item
Take $s=\frac{1}{2}$, $N=14$ and $\ell=7$.
Then the following partitions satisfy the condition ({\bf C});
$(6,1)$, $(5,2)$, $(4,3)$, $(4,1,1,1),$ $(3,2,1,1)$, $(2,2,2,1)$ and $(2,1^5)$.
From them we can construct the following 15 flip invariant rigged configurations.
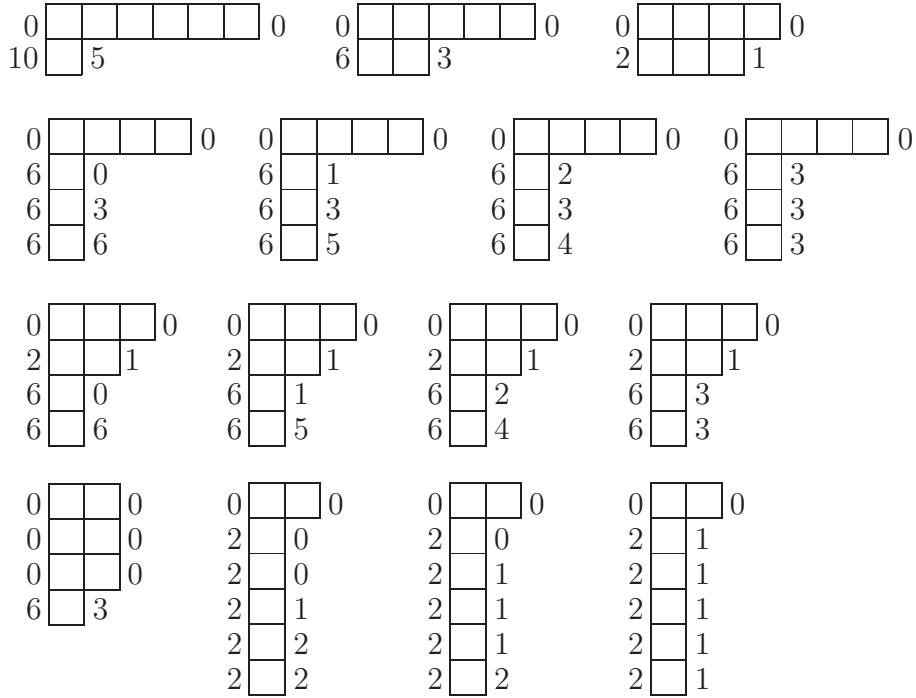
\begin{center}
\unitlength 12pt
\begin{picture}(9.5,2.0)(-0.7,0)
\put(-0.7,1.2){$0$}
\put(-1.2,0.2){$10$}
\put(0,0){\yng(6,1)}
\put(7.1,1.2){$0$}
\put(1.4,0.2){$5$}
\end{picture}
\begin{picture}(8.5,2.0)(-0.7,0)
\put(-0.7,1.2){$0$}
\put(-0.7,0.2){$6$}
\put(0,0){\yng(5,2)}
\put(5.9,1.2){$0$}
\put(2.5,0.2){$3$}
\end{picture}
\begin{picture}(10.5,2.0)(-0.7,0)
\put(-0.7,1.2){$0$}
\put(-0.7,0.2){$2$}
\put(0,0){\yng(4,3)}
\put(4.8,1.2){$0$}
\put(3.6,0.2){$1$}
\end{picture}
\end{center}

\begin{center}
\unitlength 12pt
\begin{picture}(7.0,4.5)(-0.7,0)
\put(-0.7,3.5){$0$}
\put(-0.7,2.4){$6$}
\put(-0.7,1.3){$6$}
\put(-0.7,0.2){$6$}
\put(0,0){\yng(4,1,1,1)}
\put(4.8,3.5){$0$}
\put(1.4,2.4){$0$}
\put(1.4,1.3){$3$}
\put(1.4,0.2){$6$}
\end{picture}
\begin{picture}(7.0,4.5)(-0.7,0)
\put(-0.7,3.5){$0$}
\put(-0.7,2.4){$6$}
\put(-0.7,1.3){$6$}
\put(-0.7,0.2){$6$}
\put(0,0){\yng(4,1,1,1)}
\put(4.8,3.5){$0$}
\put(1.4,2.4){$1$}
\put(1.4,1.3){$3$}
\put(1.4,0.2){$5$}
\end{picture}
\begin{picture}(7.0,4.5)(-0.7,0)
\put(-0.7,3.5){$0$}
\put(-0.7,2.4){$6$}
\put(-0.7,1.3){$6$}
\put(-0.7,0.2){$6$}
\put(0,0){\yng(4,1,1,1)}
\put(4.8,3.5){$0$}
\put(1.4,2.4){$2$}
\put(1.4,1.3){$3$}
\put(1.4,0.2){$4$}
\end{picture}
\begin{picture}(7.0,4.5)(-0.7,0)
\put(-0.7,3.5){$0$}
\put(-0.7,2.4){$6$}
\put(-0.7,1.3){$6$}
\put(-0.7,0.2){$6$}
\put(0,0){\yng(4,1,1,1)}
\put(4.8,3.5){$0$}
\put(1.4,2.4){$3$}
\put(1.4,1.3){$3$}
\put(1.4,0.2){$3$}
\end{picture}
\end{center}

\begin{center}
\unitlength 12pt
\begin{picture}(6.0,4.5)(-0.7,0)
\put(-0.7,3.5){$0$}
\put(-0.7,2.4){$2$}
\put(-0.7,1.3){$6$}
\put(-0.7,0.2){$6$}
\put(0,0){\yng(3,2,1,1)}
\put(3.6,3.5){$0$}
\put(2.4,2.4){$1$}
\put(1.4,1.3){$0$}
\put(1.4,0.2){$6$}
\end{picture}
\begin{picture}(6.0,4.5)(-0.7,0)
\put(-0.7,3.5){$0$}
\put(-0.7,2.4){$2$}
\put(-0.7,1.3){$6$}
\put(-0.7,0.2){$6$}
\put(0,0){\yng(3,2,1,1)}
\put(3.6,3.5){$0$}
\put(2.4,2.4){$1$}
\put(1.4,1.3){$1$}
\put(1.4,0.2){$5$}
\end{picture}
\begin{picture}(6.0,4.5)(-0.7,0)
\put(-0.7,3.5){$0$}
\put(-0.7,2.4){$2$}
\put(-0.7,1.3){$6$}
\put(-0.7,0.2){$6$}
\put(0,0){\yng(3,2,1,1)}
\put(3.6,3.5){$0$}
\put(2.4,2.4){$1$}
\put(1.4,1.3){$2$}
\put(1.4,0.2){$4$}
\end{picture}
\begin{picture}(10.0,4.5)(-0.7,0)
\put(-0.7,3.5){$0$}
\put(-0.7,2.4){$2$}
\put(-0.7,1.3){$6$}
\put(-0.7,0.2){$6$}
\put(0,0){\yng(3,2,1,1)}
\put(3.6,3.5){$0$}
\put(2.4,2.4){$1$}
\put(1.4,1.3){$3$}
\put(1.4,0.2){$3$}
\end{picture}
\end{center}

\begin{center}
\unitlength 12pt
\begin{picture}(6.0,6.5)(-0.7,0)
\put(-0.7,5.7){$0$}
\put(-0.7,4.6){$0$}
\put(-0.7,3.5){$0$}
\put(-0.7,2.4){$6$}
\put(0,2.2){\yng(2,2,2,1)}
\put(2.5,5.7){$0$}
\put(2.5,4.6){$0$}
\put(2.5,3.5){$0$}
\put(1.4,2.4){$3$}
\end{picture}
\begin{picture}(6.0,6.5)(-0.7,0)
\put(-0.7,5.7){$0$}
\put(-0.7,4.6){$2$}
\put(-0.7,3.5){$2$}
\put(-0.7,2.4){$2$}
\put(-0.7,1.3){$2$}
\put(-0.7,0.2){$2$}
\put(0,0){\yng(2,1,1,1,1,1)}
\put(2.5,5.7){$0$}
\put(1.4,4.6){$0$}
\put(1.4,3.5){$0$}
\put(1.4,2.4){$1$}
\put(1.4,1.3){$2$}
\put(1.4,0.2){$2$}
\end{picture}
\begin{picture}(6.0,6.5)(-0.7,0)
\put(-0.7,5.7){$0$}
\put(-0.7,4.6){$2$}
\put(-0.7,3.5){$2$}
\put(-0.7,2.4){$2$}
\put(-0.7,1.3){$2$}
\put(-0.7,0.2){$2$}
\put(0,0){\yng(2,1,1,1,1,1)}
\put(2.5,5.7){$0$}
\put(1.4,4.6){$0$}
\put(1.4,3.5){$1$}
\put(1.4,2.4){$1$}
\put(1.4,1.3){$1$}
\put(1.4,0.2){$2$}
\end{picture}
\begin{picture}(10.0,6.5)(-0.7,0)
\put(-0.7,5.7){$0$}
\put(-0.7,4.6){$2$}
\put(-0.7,3.5){$2$}
\put(-0.7,2.4){$2$}
\put(-0.7,1.3){$2$}
\put(-0.7,0.2){$2$}
\put(0,0){\yng(2,1,1,1,1,1)}
\put(2.5,5.7){$0$}
\put(1.4,4.6){$1$}
\put(1.4,3.5){$1$}
\put(1.4,2.4){$1$}
\put(1.4,1.3){$1$}
\put(1.4,0.2){$1$}
\end{picture}
\end{center}
Since there are no forbidden riggings, our result agrees with the result of \cite{HaoNepomechieSommese}.

\item
Take $s=\frac{1}{2}$, $N=12$ and $\ell=5$.
Then the following partitions satisfy the condition ({\bf C});
$(5),$ $(3,2)$ and $(2,1,1,1)$.
Corresponding to the partition $(2,1,1,1)$ we have the following
three flip invariant rigged configurations.
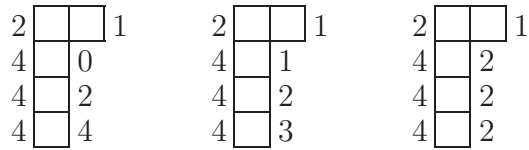
\begin{center}
\unitlength 12pt
\begin{picture}(6.0,4.5)(-0.7,0)
\put(-0.7,3.5){$2$}
\put(-0.7,2.4){$4$}
\put(-0.7,1.3){$4$}
\put(-0.7,0.2){$4$}
\put(0,0){\yng(2,1,1,1)}
\put(2.5,3.5){$1$}
\put(1.4,2.4){$0$}
\put(1.4,1.3){$2$}
\put(1.4,0.2){$4$}
\end{picture}
\begin{picture}(6.0,4.5)(-0.7,0)
\put(-0.7,3.5){$2$}
\put(-0.7,2.4){$4$}
\put(-0.7,1.3){$4$}
\put(-0.7,0.2){$4$}
\put(0,0){\yng(2,1,1,1)}
\put(2.5,3.5){$1$}
\put(1.4,2.4){$1$}
\put(1.4,1.3){$2$}
\put(1.4,0.2){$3$}
\end{picture}
\begin{picture}(6.0,4.5)(-0.7,0)
\put(-0.7,3.5){$2$}
\put(-0.7,2.4){$4$}
\put(-0.7,1.3){$4$}
\put(-0.7,0.2){$4$}
\put(0,0){\yng(2,1,1,1)}
\put(2.5,3.5){$1$}
\put(1.4,2.4){$2$}
\put(1.4,1.3){$2$}
\put(1.4,0.2){$2$}
\end{picture}
\end{center}
Since $P_1(\nu)=4$, the rigging $J_{1,1}=J_{1,2}=J_{1,3}=2$ is forbidden.
Therefore totally one has $1+1+(3-1)=4$ for the number of flip invariant
rigged configurations satisfying the condition given in ({\bf C}).

More generally, if $s=\frac{1}{2}$, $N$ is even and $\ell=5$
our conjecture predicts that the number of physical singular solutions should be
\begin{align*}
\begin{array}{ll}
\frac{N-2}{2}&\mbox{if }N\equiv 2\pmod 4,\\
\frac{N-4}{2}&\mbox{if }N\equiv 0\pmod 4.\rule{0pt}{15pt}
\end{array}
\end{align*}

\item
Take $s=3/2$, even integer $N\geq 8$ and $\ell=10$.
Then the number of rigged configurations satisfying conditions of Conjecture \ref{conj:main}
({\bf B-a}) is equal to
\begin{align*}
\frac{N-4}{N+2}{\frac{N+6}{2}\choose 3}+2.
\end{align*}
For $N=8$ this number is equal to $16$, cf. \cite{HNS}, Table 3.

\item
Take $s=3/2$, $N=8$ and $\ell=12$.
Then the following partitions satisfy the condition ({\bf B-a});
$(12)$, $(10,2)$, $(10,1,1)$, $(8,3,1)$, $(8,2,2)$,
$(8,2,1,1)$, $(7,4,1)$, $(6,5,1)$, $(6,3,3)$,
$(6,3,2,1)$, $(6,2,2,2)$, $(5,4,3)$, $(5,4,2,1)$, $(4,4,4)$ and $(4,3,3,2)$.
Then the number of flip invariant rigged configurations
corresponding to these partitions is
$$(1+1+2+1+3+1+1+4+1+1+1+1+1+1+2) = 22.$$
More generally, if $s= 3/2$, $N \ge 8$ is even and $\ell =12$, the number of 
rigged configurations satisfying  conditions of
Conjecture \ref{conj:main}, ({\bf B-a}) is equal to
$$ \frac{N-6}{N+2}~{\frac{N+8}{2} \choose 4} + 8.$$

\item
Take $s=1$, even integer $N \ge 8$ and $\ell =7$.
Then the following partitions satisfy the condition ({\bf B-b});
$(7)$, $(5,2)$, $(5,1,1)$, $(4,3)$, and $(3,2,2)$.
The number of flip invariant rigged configurations is
$$ \frac{(N-2)(N+4)}{8} .$$
For $N=8$ this number is equal to $9$, cf \cite{HNS}, Table 1.

\item
Take $s=1$, even integer $N\geq 10$ and $\ell=9$.
Then the number of rigged configurations satisfying conditions of
Conjecture \ref{conj:main} ({\bf B-b}) is equal to
\begin{align*}
\frac{N-4}{N+2}{\frac{N+6}{2}\choose 3}+2-\frac{N-2}{2}.
\end{align*}
\qed
\end{itemize}
\end{example}

\begin{corollary}
\label{cor:of_conj}
${}$

\begin{itemize}
\item
Suppose that $N$ and $\ell$ are both even.
The number of physical singular solutions to the Bethe ansatz equations
$\mathrm{BAE}(N,\ell)$ for the homogeneous spin $s$ Heisenberg chain is equal to
\begin{align}
\sum_{\nu}
\prod_{k}
{[\frac{m_k}{2}]+\frac{P_k(\nu)}{2} \choose [\frac{m_k}{2}]},
\end{align}
where the summation runs over either 
\begin{itemize}
\item
the set of partitions $\nu$ which satisfies the
condition $(${\bf B-a}$)$ of Conjecture \ref{conj:main}, if $2s$ is {\bf odd}, or
\item
the set of partitions $\nu$ which satisfies the  condition $(${\bf B-b}$)$ of 
Conjecture \ref{conj:main}, if $2s$ is {\bf even};
\end{itemize}
and for any real number $x$ the symbol $[x]$ means the {\bf  integer part} of $x$, i.e.
is a unique integer n such that $ n \le x < n+1$;

the symbol ${n+m \choose m} = \frac{(n+m) !}{n !~m !}$ means the
{\bf  binomial coefficient}.

\item
If $s=\frac{1}{2}$, $N$ is even, but $\ell$ is \underline{odd},
the number of physical singular solutions to $\mathrm{BAE}(N,\ell)$
for the homogeneous spin-$\frac{1}{2}$ Heisenberg chain is equal to
\begin{align}
\sum_{\nu}
\prod_{k}
\left\{
{[\frac{m_k}{2}]+\frac{P_k(\nu)}{2} \choose [\frac{m_k}{2}]}
-\chi_k(\nu)
\right\},
\end{align}
where the summation runs over the set of partitions $\nu$
which satisfies the condition $(${\bf C}$)$ of Conjecture \ref{conj:main};

$\chi_k(\nu)=1$ if $m_k(\nu)\geq 3$ is an odd integer and the vacancy number
$P_k(\nu)>0$ is divisible by 4, and $\chi_k(\nu)=0$ otherwise.
\end{itemize}
\end{corollary}
The surprising thing is that the sum $(27)$ can be computed.

\begin{proposition} ${}$

\begin{itemize}
\item
If $s=\frac{1}{2}$, $N$ and $\ell$ are both even, then the 
 number of physical singular solutions predicted by Corollary \ref{cor:of_conj}, is equal to
\begin{align}
{\frac{N-2}{2} \choose \frac{\ell -2}{2}}.
\end{align}

\item
If $s=\frac{1}{2}$, $N \equiv 2\pmod 4$ and $\ell$ is an 
\underline{odd} integer, then the 
 number of physical singular solutions predicted by Corollary \ref{cor:of_conj}, is equal to
\begin{align}
{\frac{N-2}{2} \choose \frac{\ell - 3}{2}}.
\end{align}
\qed
\end{itemize}
\end{proposition}

Note that for the spin-$\frac{1}{2}$ isotropic Heisenberg model,
the total number of the highest weight states of $\mathfrak{H}_N$ is
\begin{align}
{N\choose \ell}-{N\choose \ell-1}
\end{align}
for the prescribed value of $\ell$.
Thus the above conjecture also provides a conjecture for the total number
of regular solutions in this case.

\section{Discussion}
\label{sec:discussion}

\subsection{The case $N$ and $\ell$ are both odd}
\label{sec:N=9&l=3}

So far we have concentrated on the case when the system length $N$ is even.
For the spin-$\frac{1}{2}$ Heisenberg model, the paper \cite{HaoNepomechieSommese}
discovered one exceptional case when \underline{both} $N$ and $\ell$ are odd.
In such situation we do not have any flip invariant rigged configurations.
Nevertheless there are two physical singular solutions when $N=9$ and $\ell=3$
which corresponds to the partition $\nu=(2,1)$.
Below we give a list of 12 regular and physical singular solutions in this case.

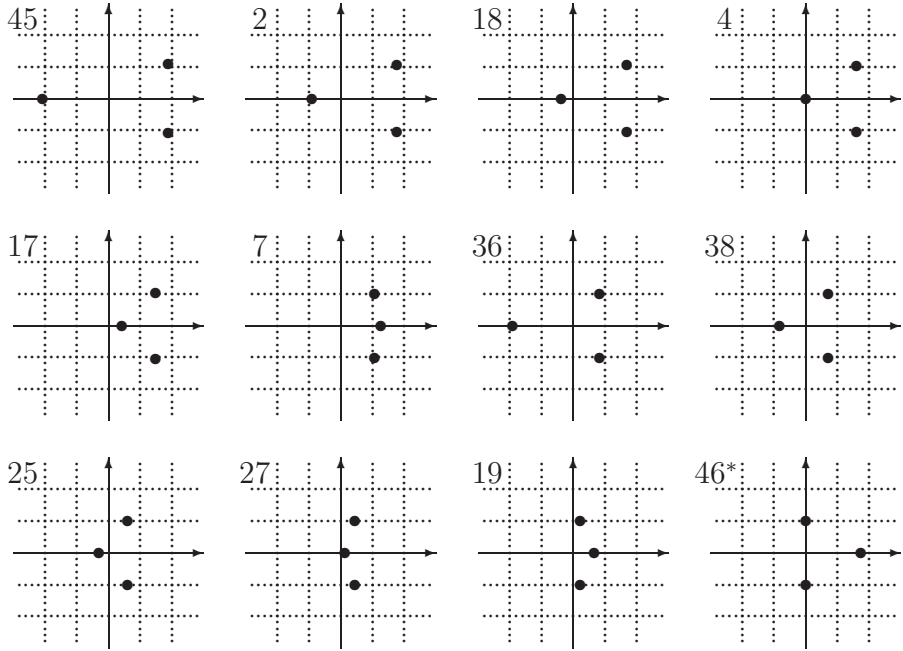
\begin{center}
\unitlength 12pt
\begin{picture}(7,6)
\put(-0.2,5.2){$45$}
\put(0,3){\vector(1,0){6}}
\put(3,0){\vector(0,1){6}}
\put(0.92,3.){\circle*{0.3}}
\put(4.88,4.08){\circle*{0.3}}
\put(4.88,1.92){\circle*{0.3}}
\multiput(1,0.21)(0,0.2){29}{\circle*{0.07}}
\multiput(2,0.21)(0,0.2){29}{\circle*{0.07}}
\multiput(4,0.21)(0,0.2){29}{\circle*{0.07}}
\multiput(5,0.21)(0,0.2){29}{\circle*{0.07}}
\multiput(0.2,1.01)(0.2,0){29}{\circle*{0.07}}
\multiput(0.2,2.01)(0.2,0){29}{\circle*{0.07}}
\multiput(0.2,4.01)(0.2,0){29}{\circle*{0.07}}
\multiput(0.2,5.01)(0.2,0){29}{\circle*{0.07}}
\end{picture}
\begin{picture}(7,6)
\put(0.2,5.2){$2$}
\put(0,3){\vector(1,0){6}}
\put(3,0){\vector(0,1){6}}
\put(2.08,3.){\circle*{0.3}}
\put(4.76,4.06){\circle*{0.3}}
\put(4.76,1.94){\circle*{0.3}}
\multiput(1,0.21)(0,0.2){29}{\circle*{0.07}}
\multiput(2,0.21)(0,0.2){29}{\circle*{0.07}}
\multiput(4,0.21)(0,0.2){29}{\circle*{0.07}}
\multiput(5,0.21)(0,0.2){29}{\circle*{0.07}}
\multiput(0.2,1.01)(0.2,0){29}{\circle*{0.07}}
\multiput(0.2,2.01)(0.2,0){29}{\circle*{0.07}}
\multiput(0.2,4.01)(0.2,0){29}{\circle*{0.07}}
\multiput(0.2,5.01)(0.2,0){29}{\circle*{0.07}}
\end{picture}
\begin{picture}(7,6)
\put(-0.2,5.2){$18$}
\put(0,3){\vector(1,0){6}}
\put(3,0){\vector(0,1){6}}
\put(2.62,3.){\circle*{0.3}}
\put(4.68,4.06){\circle*{0.3}}
\put(4.68,1.94){\circle*{0.3}}
\multiput(1,0.21)(0,0.2){29}{\circle*{0.07}}
\multiput(2,0.21)(0,0.2){29}{\circle*{0.07}}
\multiput(4,0.21)(0,0.2){29}{\circle*{0.07}}
\multiput(5,0.21)(0,0.2){29}{\circle*{0.07}}
\multiput(0.2,1.01)(0.2,0){29}{\circle*{0.07}}
\multiput(0.2,2.01)(0.2,0){29}{\circle*{0.07}}
\multiput(0.2,4.01)(0.2,0){29}{\circle*{0.07}}
\multiput(0.2,5.01)(0.2,0){29}{\circle*{0.07}}
\end{picture}
\begin{picture}(7,6)
\put(0.2,5.2){$4$}
\put(0,3){\vector(1,0){6}}
\put(3,0){\vector(0,1){6}}
\put(3.,3.){\circle*{0.3}}
\put(4.6,4.04){\circle*{0.3}}
\put(4.6,1.96){\circle*{0.3}}
\multiput(1,0.21)(0,0.2){29}{\circle*{0.07}}
\multiput(2,0.21)(0,0.2){29}{\circle*{0.07}}
\multiput(4,0.21)(0,0.2){29}{\circle*{0.07}}
\multiput(5,0.21)(0,0.2){29}{\circle*{0.07}}
\multiput(0.2,1.01)(0.2,0){29}{\circle*{0.07}}
\multiput(0.2,2.01)(0.2,0){29}{\circle*{0.07}}
\multiput(0.2,4.01)(0.2,0){29}{\circle*{0.07}}
\multiput(0.2,5.01)(0.2,0){29}{\circle*{0.07}}
\end{picture}
\end{center}

\begin{center}
\unitlength 12pt
\begin{picture}(7,6)
\put(-0.2,5.2){$17$}
\put(0,3){\vector(1,0){6}}
\put(3,0){\vector(0,1){6}}
\put(3.42,3.){\circle*{0.3}}
\put(4.48,4.04){\circle*{0.3}}
\put(4.48,1.96){\circle*{0.3}}
\multiput(1,0.21)(0,0.2){29}{\circle*{0.07}}
\multiput(2,0.21)(0,0.2){29}{\circle*{0.07}}
\multiput(4,0.21)(0,0.2){29}{\circle*{0.07}}
\multiput(5,0.21)(0,0.2){29}{\circle*{0.07}}
\multiput(0.2,1.01)(0.2,0){29}{\circle*{0.07}}
\multiput(0.2,2.01)(0.2,0){29}{\circle*{0.07}}
\multiput(0.2,4.01)(0.2,0){29}{\circle*{0.07}}
\multiput(0.2,5.01)(0.2,0){29}{\circle*{0.07}}
\end{picture}
\begin{picture}(7,6)
\put(0.2,5.2){$7$}
\put(0,3){\vector(1,0){6}}
\put(3,0){\vector(0,1){6}}
\put(4.26,3.){\circle*{0.3}}
\put(4.06,4.){\circle*{0.3}}
\put(4.06,2.){\circle*{0.3}}
\multiput(1,0.21)(0,0.2){29}{\circle*{0.07}}
\multiput(2,0.21)(0,0.2){29}{\circle*{0.07}}
\multiput(4,0.21)(0,0.2){29}{\circle*{0.07}}
\multiput(5,0.21)(0,0.2){29}{\circle*{0.07}}
\multiput(0.2,1.01)(0.2,0){29}{\circle*{0.07}}
\multiput(0.2,2.01)(0.2,0){29}{\circle*{0.07}}
\multiput(0.2,4.01)(0.2,0){29}{\circle*{0.07}}
\multiput(0.2,5.01)(0.2,0){29}{\circle*{0.07}}
\end{picture}
\begin{picture}(7,6)
\put(-0.2,5.2){$36$}
\put(0,3){\vector(1,0){6}}
\put(3,0){\vector(0,1){6}}
\put(1.08,3.){\circle*{0.3}}
\put(3.82,4.){\circle*{0.3}}
\put(3.82,2.){\circle*{0.3}}
\multiput(1,0.21)(0,0.2){29}{\circle*{0.07}}
\multiput(2,0.21)(0,0.2){29}{\circle*{0.07}}
\multiput(4,0.21)(0,0.2){29}{\circle*{0.07}}
\multiput(5,0.21)(0,0.2){29}{\circle*{0.07}}
\multiput(0.2,1.01)(0.2,0){29}{\circle*{0.07}}
\multiput(0.2,2.01)(0.2,0){29}{\circle*{0.07}}
\multiput(0.2,4.01)(0.2,0){29}{\circle*{0.07}}
\multiput(0.2,5.01)(0.2,0){29}{\circle*{0.07}}
\end{picture}
\begin{picture}(7,6)
\put(-0.2,5.2){$38$}
\put(0,3){\vector(1,0){6}}
\put(3,0){\vector(0,1){6}}
\put(2.18,3.){\circle*{0.3}}
\put(3.7,4.){\circle*{0.3}}
\put(3.7,2.){\circle*{0.3}}
\multiput(1,0.21)(0,0.2){29}{\circle*{0.07}}
\multiput(2,0.21)(0,0.2){29}{\circle*{0.07}}
\multiput(4,0.21)(0,0.2){29}{\circle*{0.07}}
\multiput(5,0.21)(0,0.2){29}{\circle*{0.07}}
\multiput(0.2,1.01)(0.2,0){29}{\circle*{0.07}}
\multiput(0.2,2.01)(0.2,0){29}{\circle*{0.07}}
\multiput(0.2,4.01)(0.2,0){29}{\circle*{0.07}}
\multiput(0.2,5.01)(0.2,0){29}{\circle*{0.07}}
\end{picture}
\end{center}

\begin{center}
\unitlength 12pt
\begin{picture}(7,6)
\put(-0.2,5.2){$25$}
\put(0,3){\vector(1,0){6}}
\put(3,0){\vector(0,1){6}}
\put(2.7,3.){\circle*{0.3}}
\put(3.6,4.){\circle*{0.3}}
\put(3.6,2.){\circle*{0.3}}
\multiput(1,0.21)(0,0.2){29}{\circle*{0.07}}
\multiput(2,0.21)(0,0.2){29}{\circle*{0.07}}
\multiput(4,0.21)(0,0.2){29}{\circle*{0.07}}
\multiput(5,0.21)(0,0.2){29}{\circle*{0.07}}
\multiput(0.2,1.01)(0.2,0){29}{\circle*{0.07}}
\multiput(0.2,2.01)(0.2,0){29}{\circle*{0.07}}
\multiput(0.2,4.01)(0.2,0){29}{\circle*{0.07}}
\multiput(0.2,5.01)(0.2,0){29}{\circle*{0.07}}
\end{picture}
\begin{picture}(7,6)
\put(-0.2,5.2){$27$}
\put(0,3){\vector(1,0){6}}
\put(3,0){\vector(0,1){6}}
\put(3.12,3.){\circle*{0.3}}
\put(3.44,4.){\circle*{0.3}}
\put(3.44,2.){\circle*{0.3}}
\multiput(1,0.21)(0,0.2){29}{\circle*{0.07}}
\multiput(2,0.21)(0,0.2){29}{\circle*{0.07}}
\multiput(4,0.21)(0,0.2){29}{\circle*{0.07}}
\multiput(5,0.21)(0,0.2){29}{\circle*{0.07}}
\multiput(0.2,1.01)(0.2,0){29}{\circle*{0.07}}
\multiput(0.2,2.01)(0.2,0){29}{\circle*{0.07}}
\multiput(0.2,4.01)(0.2,0){29}{\circle*{0.07}}
\multiput(0.2,5.01)(0.2,0){29}{\circle*{0.07}}
\end{picture}
\begin{picture}(7,6)
\put(-0.2,5.2){$19$}
\put(0,3){\vector(1,0){6}}
\put(3,0){\vector(0,1){6}}
\put(3.66,3.){\circle*{0.3}}
\put(3.22,4.){\circle*{0.3}}
\put(3.22,2.){\circle*{0.3}}
\multiput(1,0.21)(0,0.2){29}{\circle*{0.07}}
\multiput(2,0.21)(0,0.2){29}{\circle*{0.07}}
\multiput(4,0.21)(0,0.2){29}{\circle*{0.07}}
\multiput(5,0.21)(0,0.2){29}{\circle*{0.07}}
\multiput(0.2,1.01)(0.2,0){29}{\circle*{0.07}}
\multiput(0.2,2.01)(0.2,0){29}{\circle*{0.07}}
\multiput(0.2,4.01)(0.2,0){29}{\circle*{0.07}}
\multiput(0.2,5.01)(0.2,0){29}{\circle*{0.07}}
\end{picture}
\begin{picture}(7,6)
\put(-0.5,5.2){$46^\ast$}
\put(0,3){\vector(1,0){6}}
\put(3,0){\vector(0,1){6}}
\put(4.74,3.){\circle*{0.3}}
\put(3.,4.){\circle*{0.3}}
\put(3.,2.){\circle*{0.3}}
\multiput(1,0.21)(0,0.2){29}{\circle*{0.07}}
\multiput(2,0.21)(0,0.2){29}{\circle*{0.07}}
\multiput(4,0.21)(0,0.2){29}{\circle*{0.07}}
\multiput(5,0.21)(0,0.2){29}{\circle*{0.07}}
\multiput(0.2,1.01)(0.2,0){29}{\circle*{0.07}}
\multiput(0.2,2.01)(0.2,0){29}{\circle*{0.07}}
\multiput(0.2,4.01)(0.2,0){29}{\circle*{0.07}}
\multiput(0.2,5.01)(0.2,0){29}{\circle*{0.07}}
\end{picture}
\end{center}
Here some remarks are in order;
\begin{itemize}
\item
the above solutions are arranged according to the real parts of length 2 strings of solutions,
\item
the spacing of dotted lines is $0.5$,
\item
the label of each solution corresponds to the label in
the supplementary table of \cite{HaoNepomechieSommese}
(table $N=9$, $M=3$);
label with asterisk ($46^*$ in this case) means that the solution
is singular and physical,
\item
the remaining 12 solutions are obtained by multiplying $(-1)$ to
each root in the above 12 solutions.
\end{itemize}

For each solution in the above table,
it is natural to associate the following rigged configurations.
The first six solutions correspond to
\begin{center}
\unitlength 12pt
\begin{picture}(4,2.5)(-0.7,0)
\put(-0.7,1.2){$3$}
\put(-0.7,0.2){$5$}
\put(0,0){\yng(2,1)}
\put(2.5,1.2){$3$}
\put(1.4,0.2){$r$}
\end{picture}
\end{center}
where $r=0,\ldots,5$ according to the order of the above table.
The next six solutions correspond to
\begin{center}
\unitlength 12pt
\begin{picture}(4,2.5)(-0.7,0)
\put(-0.7,1.2){$3$}
\put(-0.7,0.2){$5$}
\put(0,0){\yng(2,1)}
\put(2.5,1.2){$2$}
\put(1.4,0.2){$r$}
\end{picture}
\end{center}
where $r=0,\ldots,5$ according to the order of the above table.

To summarize, the exceptional physical singular solutions
in the case of $N=9$ and $\ell=3$ correspond to the following rigged configurations;
\begin{center}
\unitlength 12pt
\begin{picture}(6,2.5)(-0.7,0)
\put(-0.7,1.2){$3$}
\put(-0.7,0.2){$5$}
\put(0,0){\yng(2,1)}
\put(2.5,1.2){$2$}
\put(1.4,0.2){$5$}
\end{picture}
\begin{picture}(4,2.5)(-0.7,0)
\put(-0.7,1.2){$3$}
\put(-0.7,0.2){$5$}
\put(0,0){\yng(2,1)}
\put(2.5,1.2){$1$}
\put(1.4,0.2){$0$}
\end{picture}
\end{center}
It will be an interesting problem to find general rule
to characterize the rigged configurations corresponding
to physical singular solutions
for the case when both $N$ and $\ell$ are odd.

\subsection{On the number of solutions to the Bethe ansatz equations}

Follow \cite{HaoNepomechieSommese},
let us denote by ${\cal{N}}(N,\ell)$ the number of solutions with 
pairwise $\ell$ distinct roots, and by ${\cal{N}}_{sp}(N,\ell)$ the number of physical singular solutions  
to the Bethe ansatz equations for the spin-$\frac{1}{2}$ Heisenberg model of 
length $N$.

\begin{conjecture}

\label{conj:number_of_roots}
${}$

\begin{itemize}
\item
Assume that $N$ and $\ell > 3$ are both even, then
\begin{align}
{\cal{N}}(N,\ell) + {\cal{N}}_{sp}(N,\ell) = {N-1 \choose \ell},\qquad 
\mbox{if }\quad 2\ell \le N.
\end{align}
In other words, 
\begin{align}
{\cal{N}}(N,\ell) = { N-1 \choose \ell} -{\frac{N-2}{2} \choose \frac{\ell- 2}{2} }.
\end{align}

\item
Assume that $N \equiv 2 \pmod 4$, $ N \ge 6$,  but  $\ell \ge 3$ is odd, then
\begin{align}
{\cal{N}}(N,\ell) + {\cal{N}}_{sp}(N,\ell) = {N-1 \choose \ell},\qquad \mbox{if }
\quad 2\ell \le N.
\end{align}
In other words, 
\begin{align}
{\cal{N}}(N,\ell) = { N-1 \choose \ell} -{\frac{N-2}{2} \choose \frac{\ell- 3}{2} }.
\end{align}

\item
If $N \ge 3$ is odd, but $\ell \ge 4$ is even, then
\begin{align}
{\cal{N}}(N,\ell) + {\cal{N}}_{sp}(N-1,\ell -2) = {N-1 \choose \ell},\qquad
\mbox{if}\quad 2\ell \le N.
\end{align}
In other words,
\begin{align}
{\cal{N}}(N,\ell) = { N-1 \choose \ell} -{\frac{N-3}{2} \choose \frac{\ell - 4}{2} },\quad
{\cal{N}}_{s}(N,\ell)= {N-1 \choose \ell -2} -{\frac{N-3}{2} 
\choose \frac{\ell - 4}{2} }. 
\end{align}

\item
If $N \ge 3$ and $\ell \ge 3$  are both odd, then
\begin{align}
{\cal{N}}(N,\ell) + {\cal{N}}_{sp}(N,\ell) = { N-1 \choose \ell},\quad
{\cal{N}}_{s}(N,\ell) ={N-1 \choose \ell -2}.
\end{align}
\qed
\end{itemize}
\end{conjecture}
Therefore, Conjecture \ref{conj:number_of_roots}
predicts that if $\ell \ge 2$  and $N \ge 2$ have the same \underline{parity}, 
then the number of singular solutions 
${\cal{N}}_{s}(N,\ell)$ to the Bethe equations in question, is equal to
\begin{align}
{\cal{N}}_{s}(N,\ell) ={N-1 \choose \ell -2}.
\end{align}
For example,
$${\cal{N}}(14,6)+{\cal{N}}_{sp}(14,6)=1716={13 \choose 6},~~{\cal{N}}_{sp}(14,6) =15 ={6 \choose 2},~~
{\cal{N}}_{s}(14,6)= 715={13 \choose 4},$$
$${\cal{N}}(14,5)+ {\cal{N}}_{sp}(14,5) = 1287= {13 \choose 5},~~{\cal{N}}_{sp}(14,5) = 6 ={6 \choose 1},~~
{\cal{N}}_{s}(14,5) =286= {13 \choose 3} ,$$
$$ {\cal{N}}(13,6)+ {\cal{N}}_{sp}(12,4) = 919+5= {12 \choose 6},~~~ 
{\cal{N}}_{s}(13,6) = 490 = {12 \choose 4} -{5 \choose 1},$$
$$ {\cal{N}}(9,3) + {\cal{N}}_{sp}(9,3) = 54+2={8 \choose 3},~~
{\cal{N}}_{s}(9,3)= 8 ={8 \choose 1}.$$
\underline{However}, 
$${\cal{N}}(12,5)+ {\cal{N}}_{sp}(12,5) = 456+4= 460 < {11 \choose 5} = 462,~
{\cal{N}}_{s}(12,5) =163 < {11 \choose 3} =165.$$

\subsection{On Conjecture \ref{conj:main} $({\bf C})$}

Probably, if $N \equiv 0 \pmod 2$ and  $\ell$ is an odd number, then  for
 $k$  such that $m_k \ge 3$ and odd, and 
$P_k(\nu) \equiv 0~ \pmod 4$ and  $P_{k}(\nu) > 0,$  it is more natural to 
allow only   riggings with strict inequalities:
$$ 0 \le J_{k,1} < J_{k,2} < \ldots < J_{k.m_{k}} \le \frac{P_k(\nu)}{2}.$$
For example, if $\ell=7$, the number of such rigged configurations is equal to 
$$(N-2)(N-4)/8 - N +9,$$ whereas the number of 
rigged configurations which satisfies conditions of Conjecture \ref{conj:main}
$({\bf C})$ is equal to 
$$(N-2)(N-4)/8 -3.$$

\subsection{Some related topics from mathematical physics}
It should be worth while to mention that the theory of the rigged configurations
is extensively studied from various points of view.
Indeed, it is widely believed that the rigged configurations exist for
finite dimensional representations of arbitrary quantum affine algebras.
Especially, we would like to mention that there is a clear physical
interpretation of the rigged configurations.
See \cite{S:review} for an introductory review related with spin-$\frac{1}{2}$ case
of $\mathfrak{sl}_2$ which is the main case of the present paper.

The main point of the rigged configuration theory
is the bijection between the rigged configurations
and the tensor products of crystals.
In the spin-$\frac{1}{2}$ case of $\mathfrak{sl}_2$,
one can regard the latter objects as sequences of the letters 1 and 2
which we call crystal paths.
On the crystal paths we can define a discrete soliton system called
the box-ball system \cite{TS,Tak}.
Then the fundamental observation of \cite{KOSTY} is that
the rigged configurations provide a complete set of the action and angle variables
of the box-ball systems.
More precisely, each row of the configuration $\nu$ corresponds to a soliton
whose position is specified by the corresponding riggings.
This soliton picture is also confirmed from the point of view of
the ordinary soliton theory (the KP equation) \cite{KSY}.

Finally, we would like to mention that in the spin-$\frac{1}{2}$ case of $\mathfrak{sl}_2$,
there is a periodic version of the box-ball systems \cite{YuraTokihiro}
which admit the rigged configuration approach \cite{KTT}.
By taking a suitable limit of initial value solutions of the linear box-ball systems \cite{KSY,Sak1},
we can solve the initial value problem for the periodic case in terms of
the tropical Riemann theta functions \cite{KS}.
This is a direct discrete analogue of the periodic solution for the KP equation
obtained by B.~A.~Dubrovin, V.~B.~Matveev and S.~P.~Novikov \cite{DMN}.

\subsection{Some related combinatorics}
Finally, let us mention some related topics from the point of view of pure 
combinatorics.  Here we consider only the case of ${\mathfrak{gl}}(N)$ case.

The starting data for definition of rigged configurations are:~~\\
a partition $\lambda =(\lambda_1,\ldots, \lambda_N \ge 0)$, ~~~and  \\
a collection of rectangular shape partitions 
$$R:= \{ R_a= (\underbrace{\mu_a,\ldots, \mu_a}_{\eta_{a}}) \}~~
such~~~ that~~~ \sum_{j \ge 1}~\lambda_j = \sum_{a} ~\mu_a~\eta_a. $$

These data come from the analyses of the the Bethe ansatz equations 
corresponding to the ${\mathfrak{gl}}(N)$ ~ $XXX$ model of ``spin''   $R$, ~
based on the use of the  so-called {\it String Conjecture}, see e.g. 
\cite{Kir}.

The main results concerning the rigged configuration theory discovered in 
\cite{KirillovReshetikhin}, \cite{Kir2}, \cite{Kir4}, \cite{KSS:2002}  are

$\bullet$~~~The number of rigged configuration related with pair $(\lambda, R)$ is equal to the tensor product 
multiplicity 
\begin{align}
\label{eq:kirillov}
 {\rm Mult} [ V_{\lambda}^{{\mathfrak{gl}}(N)} : \bigotimes_{a \ge 1}~
V_{R_{a}}^{{\mathfrak{gl}}(N)} ],
\end{align}
where $V_{\mu}^{{\mathfrak{gl}}(N)}$~~stands for the irreducible 
representation of the Lie algebra ${\mathfrak{gl}}(N)$ corresponding to 
partition $\lambda$.

$\bullet$~~~There exists a bijection, called {\it Rigged Configuration 
Bijection},~($RC$-bijection  for short), ~between the set of rigged 
configurations related with pair $(\lambda,R)$,~and the set of so-called 
{\it Littlewood--Richardson tableaux}~ which are some combinatorial objects 
describing the tensor product multiplicity introduced above. 
 
The $RC$-bijection has a big parity of deep and sometimes unexpected properties related with
Algebraic Combinatorics \cite{Kir2},  Representation Theory 
\cite{Kir3}, \cite{KOSTY}, Integrable 
Systems \cite{KS}, \cite{KSY}, \cite{KTT}  and {\it etc}.~In the present paper 
we state only one unexpected  
(at least for A.N.K)  result discovered by the first author, 
see \cite{Kir1}, \cite{KSS:2002} for proofs, 
 that the flip map on the set of rigged configurations related with pair $(\lambda,R)$, corresponds to the 
so-called  {\it $Sch\ddot{u}tzenberger$ involution} on the set of Littlewood--Richardson tableaux needed to 
 describe the tensor product multiplicity (\ref{eq:kirillov}). 

In the case of ${\mathfrak{sl}}_2$ spin-$\frac{1}{2}$ Heisenberg model, the $RC$-bijection gives rise to a 
bijection between the set of rigged configurations $RC(\ell, N)$ and the set of standard Young tableaux of the 
shape $(N-\ell,\ell)$.~~It is well-known that if $N$ is even, then the set of standard Young tableaux of 
shape $(N-\ell, \ell)$ which are invariant under the action of the  Sch\"{u}tzenberger involution, is 
in one-to-one correspondence with the set of standard {\it domino tableaux} of the same shape, see e.g. 
\cite{CL}, \cite{BK}.
Our Conjecture \ref{conj:main} holds that the set of singular 
physical solutions to the $BAE$  is in a bijection with a set of special 
domino tableaux depending on a ${\mathfrak{gl}}(2)$-$XXX$ mode chosen. We 
expect a similar connection in general case.

\paragraph{Acknowledgments:}
The work of RS is partially supported by
Grants-in-Aid for Scientific Research No.25800026 from JSPS.

\end{document}